\documentclass{pasj00}
\draft
\usepackage{pdflscape}
\usepackage{rotating}
\usepackage{float}
\usepackage{subfig}
\usepackage{graphicx}
\usepackage[normalem]{ulem}
\usepackage{url}
\usepackage{lscape}
\usepackage{color}

\newcommand{\chg}[1]{#1}

\begin{document}
\SetRunningHead{Castro et al.}{{\sl AKARI} Observation of AGNs: PAH 3.3 $\micron$ Feature}
\title{{\sl AKARI} Infrared Camera Observations of the 3.3 $\micron$ PAH Feature in {\sl Swift}/BAT AGNs}

\author{Angel \textsc{Castro}\altaffilmark{1},Takamitsu \textsc{Miyaji}\altaffilmark{1,2},
  Mai \textsc{Shirahata}\altaffilmark{3,4}, Kohei \textsc{Ichikawa}\altaffilmark{5}, Shinki \textsc{Oyabu}\altaffilmark{6},
  David M. \textsc{Clark}\altaffilmark{1}, Masatoshi \textsc{Imanishi}\altaffilmark{7}, Takao \textsc{Nakagawa}\altaffilmark{4},
  Yoshihiro \textsc{Ueda}\altaffilmark{5}}
\altaffiltext{1}{Instituto de Astronom\'ia, Universidad Nacional
   Aut\'onoma de M\'exico (UNAM), Ensenada, Baja California, M\'exico 
   {\center (mailing address: PO Box 439027, San Diego, CA 92143-9027, USA)}}
\altaffiltext{2}{University of California, San Diego, Center for Astrophysics and Space Sciences, 9500 Gilman Drive, 
  La Jolla, CA 92093-0424, USA}
\altaffiltext{3}{National Astronomical Observatory of Japan (NAOJ), 2-21-1 Osawa, Mitaka, Tokyo 181-8588, Japan}
\altaffiltext{4}{Institute of Space and Astronautical Science (ISAS), Japan Aerospace Exploration Agency (JAXA),
3-1-1 Yoshino-dai, Chuo-ku, Sagamihara 252-5210}
\altaffiltext{5}{Department of Astronomy, Kyoto University, Kitashirakawa-Oiwake-cho, Sakyo-ku, Kyoto 606-8502}
\altaffiltext{6}{Graduate School of Science, Nagoya University, Furo-cho, Chikusa-ku, Nagoya, Aichi 464-8602, Japan}
\altaffiltext{7}{Subaru Telescope, 650 North A’ohoku Place, Hilo, Hawaii, 96720, USA}

\email{acastro@astrosen.unam.mx}

\KeyWords{galaxies: active ---  galaxies: Seyfert --- X-rays: galaxies}
\maketitle

\begin{abstract}
We explore the relationships between the 3.3 $\micron$ polycyclic aromatic hydrocarbon (PAH) feature and active galactic nucleus (AGN)
properties of a sample of 54 hard X-­ray selected bright AGNs, including both Seyfert 1 and Seyfert 2 type objects, using
the InfraRed Camera (IRC) on board the infrared astronomical satellite {\sl AKARI}. The sample is selected from the 9-­month {\sl
Swift}/BAT survey  in the 14--­195 keV band and all of them have measured X-­ray spectra at $E \lesssim 10$ keV.  These X-­ray
spectra provide measurements of the neutral hydrogen column density ($N_{\rm H}$) towards the AGNs.  We use the 3.3 $\micron$ PAH
luminosity  ($L_{\rm 3.3{\micron}}$) as a proxy for star formation activity and hard X-ray luminosity ($L_{\rm 14-195keV}$)
as an indicator of the AGN activity. We search for possible difference of star-­formation activity between type 1 (un-absorbed)
and type 2 (absorbed) AGNs. We have made several statistical analyses taking the upper-limits of the PAH lines into account
utilizing survival analysis methods. The results of our $\log(L_{\rm 14-195keV})$ versus $\log(L_{\rm 3.3{\micron}})$ regression shows a
positive correlation and the \chg{slope for the type 1/unobscured AGNs is steeper than that of type 2/obscured AGNs at a 
$3\sigma$ level}. Also our analysis show that the circum-nuclear star-formation is more enhanced in type 2/absorbed AGNs than type
1/un-absorbed AGNs for low X-ray luminosity/low Eddington ratio AGNs, while there is no significant dependence of star-formation activities
on the AGN type in the high X-ray luminosities/Eddington ratios.
\end{abstract}

\section{Introduction}
\label{sec:intro}

A fundamental question on the accretion onto supermassive black holes (SMBHs) at the centers of galaxies is the fueling mechanism, where gas
is accreted from a kilo-parsec scale to a sub-parsec scale towards the black hole (BH; e.g. \cite{alexander12}; for review). Some of the
important mechanisms that can be responsible for this process are wind from the circum-nuclear star formation region (e.g.
\cite{umemura97,ohsuga01,kawakatu08}), tidal triggering by a companion galaxy (\cite{noguchi88}; for review) or a minor merger with a
satellite galaxy (\cite{gaskell85,mihos94,taniguchi97}). On the other hand, it is also suggested that once AGNs are ignited, feedback from
AGNs may clear surroundings from cold gas and quench star formation (e.g. \cite{bundy08,lagos08}). In this context, investigating star
formation activity in various types of AGNs is relevant in order to give observational clues to these scenarios.

There are some lines of observational evidence that the simplest version of unified theory of active galactic nuclei (AGNs), 
which postulates that difference between type 1 and type 2 AGNs are solely the viewing angle effect (e.g. \cite{antonucci93,urry95}), needs
modifications.  It has long been recognized that the fraction of absorbed (type 2) AGNs decrease with luminosity
\citep{lawrence-elvis82,ueda03,lafranca05,shinozaki06,hasinger08,ueda14}, although it might be because of the selection effect 
based on X-rays or optical emission lines \citep{lawrence-elvis10}. A similar trend has been observed in optical/IR (e.g.
\cite{maiolino07}) and \citet{ichikawa12b} in the X-ray/IR. \citet{simpson05} also found that the fraction of type 1 AGNs increases with
luminosity and shows that the faint-end slope of the AGN luminosity function steepens considerabily when a correction for the 'missing' type
2 is made.

 Also, clustering studies indicate some systematic difference of large-scale
environments between type-1 and type-2 AGNs \citep{cappelluti10,allevato11} (but see also \cite{hickox11} for results for 
absorbed and un-absorbed QSOs). These observations suggest that type 1 (unabsorbed) and type 2 (absorbed) AGNs have some systematic
differences in their intrinsic properties, beyond the viewing angle effect, such as opening angle/distribution of the absorbing 
material (e.g. \cite{ramos-almeida11,elitzur12}), and these two classes may be in different stages of AGN evolution, with a significant 
overlap. If a circum-nuclear  starburst plays a major role in feeding the central SMBH at the early stage of
the AGN activity, where the absorbing torus may have a thicker geometry with a larger covering factor, it is more likely to be observed as
a type 2 AGN. 

 The use of very hard X-ray ($E\gtrsim 10$ keV) surveys such as available with {\sl Swift} Burst Alert Telescope (BAT), \citep{tueller08,
tueller10,ajello12} or {\sl INTEGRAL} \citep{krivonos10} allows us to select AGNs with a wide range of absorbing column densities, since
photoelectric absorption is negligibly small.  These very hard X-ray surveys provide an efficient way of constructing a  clean and highly
unbiased census of AGNs activities in the universe, which include those that are heavily obscured up to moderately Compton-thick column
densities ($\log(N_{\rm H}){\rm [cm^{-2}]}\lesssim 25$). The {\sl Swift}/BAT AGN catalog is one of the well-studied surveys including soft
X-ray, optical, and infrared observations (e.g. \cite{winter09,winter10,ichikawa12a}).

The polycyclic aromatic hydrocarbon (PAH) features have been used to disentangle between AGN and starbursts (SB) in Ultra-Luminous
InfraRed Galaxies (ULIRGs; e.g. \cite{sanders88,lutz98}), since observationally  these features have been found to be weak or absent in
classical AGNs but generally strong in starbursts (\cite{moorwood86,genzel98,imanishi00}). The PAH emission act as an
indicator for the presence of pumping far-ultraviolet (FUV) photons and reveals the presence of massive stars \citep{genzel98,tielens08}. 
The source of the UV radiation is generally considered to be from the massive stars in the star-formation region rather than AGNs,
because in AGNs, X-ray photons destroy the PAH molecules \citep{voit92}. Thus generally PAH emission features seen in the infrared
spectra at, e.g. 3.3 \micron, 6.2 \micron, 7.7 \micron, 8.6 \micron, and  11.2 \micron, may be used as an indicator of star-formation
activity with little contamination from AGNs, and thus provides a tool for investigating star formation activities in
AGNs.

PAH molecules could be excited by UV photons from AGNs in some circumstances without being destroyed. However, such circumstances 
are limited. For example, \citet{howell07} measured bright PAH knots directly along the ionization cone of the Seyfert 2 galaxy NGC 1068. 
Even in that case, it was not clear whether the AGN radiation can directly enhance the PAH emission or it is a result of 
stimulating the formation of OB stars and UV photons from these excite PAHs. 

There have been a number of studies that have investigated the PAH emission in a sample of known AGNs and investigated the differences in
PAH emission properties among various types of AGNs. \citet{clavel00} and \citet{freudling03} showed that weak PAH and hot dust are more
associated 
with type 1 AGNs while cooler dust and strong PAHs with type 2 AGNs. \citet{haas05} argued that nuclear starburst should be weaker in
low-luminosity AGNs. Studying the stellar population of the central $\sim$200 pc of a sample of 79 nearby galaxies, most of them Seyfert 2s,
\citet{cid04} found no correlation between the star formation in the nucleus, neither for the host morphology nor for the presence of
companions. The star formation history deduced from their study varied significantly among Seyfert 2s. 

\citet{imanishi03} and \citet{imanishi04} investigated the relation between nuclear SB and AGN activities in a sample of 32 Seyfert 2 
galaxies and 23 Seyfert 1 galaxies using ground-based spectroscopy. They found that SB correlates with nuclear activity. However, they found
no significant difference between type 1 sources and type 2 sources. Similar studies by \citet{watabe08} and by \citet{oi10} found no
significant difference of star-formation between type 1 and type 2 Seyferts either.
 
\citet{diamond-stanic12} measured the AGN luminosity of a sample of Seyfert galaxies
using the [O IV]$\lambda$25.89 $\micron$ emission line and the star-forming luminosity using the 11.3 $\micron$ aromatic feature. They 
found strong correlation in the relationship between nuclear star formation ratio (SFR) (measured on r = 1 kpc scales) and the BH accretion
rate but only weakly correlated with extended (r $>$ 1 kpc) star formation in the host galaxy. Their results do not exhibit any
statistically significant differences between type 1 and type 2 Seyfert objects. 

In order to investigate the least posible biased AGN sample, we have performed 2.5--5 $\micron$ infrared spectroscopy of hard X-ray selected
AGNs from the 9-month catalog \citep{tueller08} of the  {\sl Swift}/BAT survey with the grism mode of the InfraRed Camera (IRC) instrument
on board  the Japanese space infrared observatory {\sl AKARI}. The sample contains AGNs with a wide range of $N_{\rm H}$, including both
highly obscured AGNs and unobscured AGNs. In this work, we use the 3.3 $\micron$ PAH emission detected in our spectral range as a proxy for
the star formation activity to explore the link between AGN activity, column densities towards the nucleus, AGN type and star formation.

This paper is organized as follows. In section \ref{sec:sample-selection} we explain our sample selection criteria. In section
\ref{sec:observations}, we describe the {\sl AKARI}/IRC observation. The data reduction of the IRC spectra and the subsequent
measurements of the 3.3 $\micron$ PAH flux are explained in section \ref{sec:data-reduction}. In section \ref{sec:regression-analysis} the
regression analysis and statistical tests employed are described. Results of the research are shown in
section \ref{sec:results}. Discussion and conclusions follows in sections \ref{sec:discussion} and \ref{sec:conclusions},
respectively. Throughout this paper, luminosities are calculated using $H_{0}= 75 {\rm km\, s^{-1} Mpc^{-1}},
{\Omega}_{\rm m}=0.3$ and ${\Omega}_{\rm \Lambda}=0.7$.

\section{Sample Selection} \label{sec:sample-selection}

In this work, we have selected our sample of AGNs from the 9-month {\sl Swift}/BAT catalog \citep{tueller08} for the {\sl AKARI}/IRC
spectroscopy. We have excluded blazars, where the  X-ray emission is dominated by highly collimated beams towards us. Almost all of the AGNs
have published  measurements of detailed X-ray spectroscopy at $E\lesssim 10$ [keV] from {\sl XMM-Newton}, {\sl ASCA}, {\sl Chandra}, {\sl
Beppo-SAX}, {\sl Suzaku}, and {\sl Swift} X-ray Telescopes \citep{winter09,ichikawa12a}. The most important quantity derived from the 
$E\lesssim 10$ [keV] X-ray spectroscopy is the absorbing column density of the neutral gas, expressed by the equivalent hydrogen column 
density, $N_{\rm H}$. 

Thirty-two {\sl Swift}/BAT AGNs are from our own observations made as a part of the ``AGNUL" (AGN and ULIRG) 
group proposal for the {\sl AKARI} Mission Program 3 (MP3), which covers the post-helium phase of the {\sl AKARI} mission. 
In the fist cycle of MP3, we have selected our objects among highly absorbed AGNs ($\log(N_{\rm H})>23.5 {\rm [cm^{-2}]}$) 
and well-known bright AGNs. In the second cycle, we selected our targets such that the sample is
spread over all $\log(N_{\rm H})$ levels. The remaining 22 have been observed by other groups and we obtained the data from public
archives. 
The archival data that we have used for our analysis were from observations with the same instrumental
configuration. We intended to obtain spectra of almost all remaining non-blazar AGNs in the 9-month {\sl Swift}/BAT 
catalog during the third cycle of the MP3 program. However, it became impossible due to the unfortunate 
failure of the mechanical cryogenic cooler on board {\sl AKARI}, which happened in the winter of 2010. While obtaining spectra for all the 
AGNs in a complete sample is desirable, the selection criteria of our current sample are mainly based on the X-ray absorption (first cycle)
and visibility considerations (both cycles), rather than the far infrared properties or any star-formation indicator. Also
the abstracts of the proposals of the observers of the archival data show that they did not select based on star-formation indicators. This
is in contrast with other studies that use ULIRGS/LIRGS. Thus our sample enables us to probe the star formation in AGNs in an unbiased
manner.

\begin{figure}
  \begin{center}
    \FigureFile(60mm,60mm){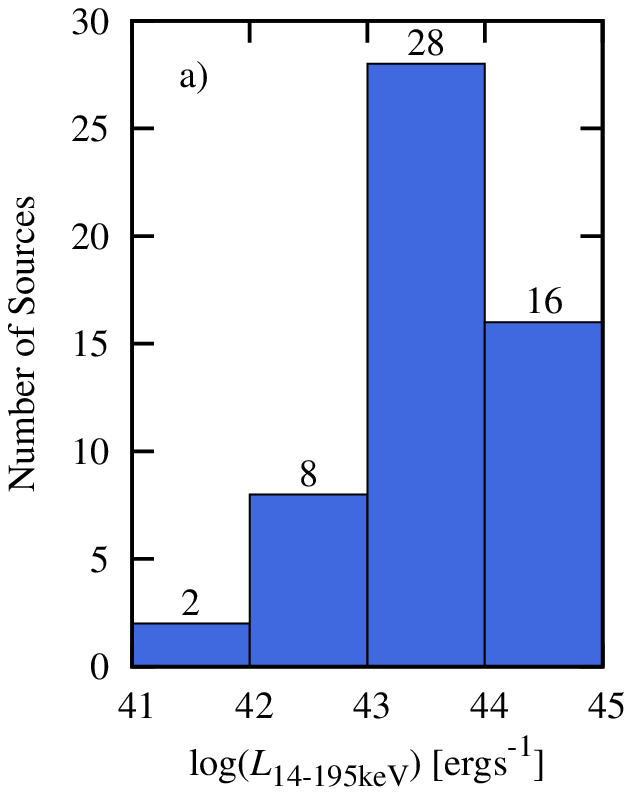}
    \FigureFile(60mm,60mm){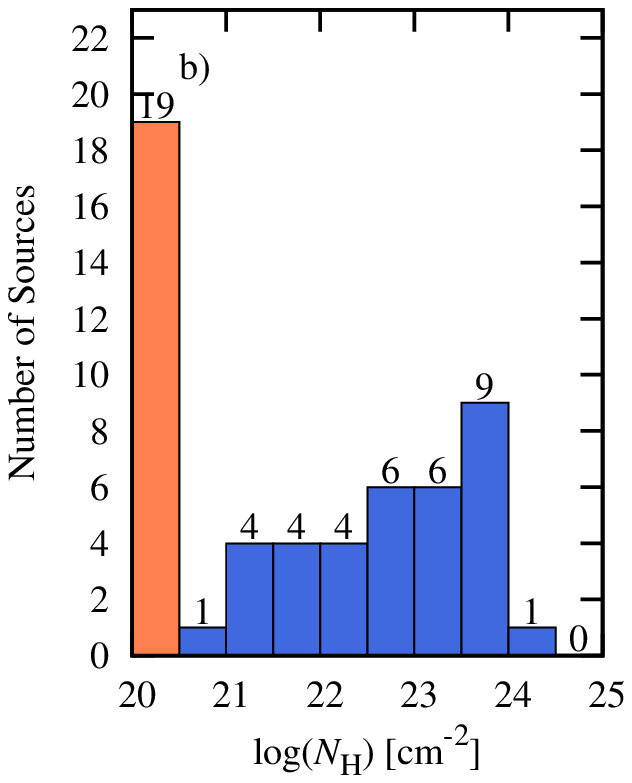}
  \end{center}
  \caption{(\emph{a}) Distribution of the $\log(L_{\rm 14-195keV})$ of the sample. (\emph{b}) Distribution of the absorbing column density
($N_{\rm H}$ in units of cm$^{-2}$) obtained from softer X-ray ($E<10$ keV) spectra of our sample (\cite{winter09,ichikawa12a}). 
(Color figures are available on the electronic version only.)}
\label{fig:sample-selection}
\end{figure}

In our sample (see table \ref{tab:obs-log}), 26 AGNs are optical type 1 AGNs (Seyfert optical type $\leq 1.5$) and 28 type 2 AGNs 
(Seyfert optical type $>1.5$). Our selected sample also has detailed X-ray spectra from the {\sl XMM-Newton}, 
{\sl Chandra, ASCA, Suzaku}, and {\sl SWIFT}/XRT in $0.3 \lesssim E [{\rm keV}]\lesssim 12$ \citep{winter09,ichikawa12a}. The
distribution of $\log(L_{\rm 14-195~keV})$ for the overall sample is shown in figure \ref{fig:sample-selection}(\emph{a}). For all objects
in our sample X-ray-derived neutral hydrogen column densities were obtained from these spectra.  Figure \ref{fig:sample-selection}(\emph{b})
shows the $N_{\rm H}$ distribution of the sample. In the cases where $N_{\rm H}$ value is not explicitly provided by \citet{winter09} (in
those cases where the original X-ray spectra were well fit by a simple absorbed power law model and thus, consistent with a un-absorbed AGN)
we took the $N_{\rm H}$ value from \citet{ichikawa12a}.

\section{Observations}\label{sec:observations}

Infrared 2.5--5 $\micron$ spectroscopy of our hard X-ray selected AGNs was performed with the IRC spectrograph \citep{ircmanual09} on
board the {\sl AKARI} Infrared satellite \citep{murakami07} during the Phase 3-mission program. The NIR channel of IRC has two dispersion
spectroscopic elements, NP (low resolution prism) and NG (high resolution grism). The spectroscopic observations can be made with or
without a slit. 

For our observations, the high resolution grism (NG) were used. Among the three slit/window sizes available, the $1^\prime\times 1^\prime$ 
square window was used for all of our observations. This window size is optimized for the spectroscopy of the point sources, where the 
size was determined such that the aperture is larger than the absolute pointing accuracy  of the satellite ($\lesssim 30^{\prime\prime}$). 
This configuration is designated as b; Np (NG for point sources; \cite{ircmanual09}). The spectral resolution of
the NG is $R=\lambda/\delta \lambda=$ 120 at $\lambda = 3.6~\micron$ \citep{ohyama07}  for point sources. The Astronomical
Observation template (AOT) of our observations was IRCZ4, where, during an orbit of observation on target, 4 spectroscopic exposures are
made with the grism, followed by a reference imaging exposure without a disperser, and 4 additional exposures with the grism. Five dark
frames are taken before and after the observations of the target.

 With the increased number of hot pixels during the phase 3 period, we followed the recommendation to make redundant observations with at 
least 3 orbits for one target. Thus, for those objects we proposed, we attempted to make 3 or 5 orbits of observation on each. 
However, not all requested observations were finally achieved and some objects have only one or two orbits of observations. The log of 
observations is shown in table \ref{tab:obs-log}.

\section{Data Reduction and Analysis} \label{sec:data-reduction}

\subsection{Reduction}
The spectra have been reduced using the IDL package, ``IRC Spectroscopy Toolkit for Phase 3 data Version 20110301"
\citep{ohyama07} \footnote{\url{http://www.ir.isas.jaxa.jp/ASTRO-F/Observation/}} (hereafter referred to as ``the toolkit''). The toolkit
performs the basic reduction pipeline of linearity correction, background and dark subtraction and the division by flat frames of the
two-dimensional (2D) spectra \citep{ircmanual09}. During the pipeline processing, the toolkit removes the hot and bad pixels upon coadding
individual images and/or upon correcting the image by its own dark image. For the NG grism, the 2D spectra corresponds to
$d\lambda=0.0097$ $\micron$/pix along the dispersion and the $1^{\prime\prime}.46$/pix perpendicular to it \citep{ohyama07}. We
adopted a narrow aperture of 3 pixels (\ttfamily{nsum=3}\rmfamily; 4$^{\prime\prime}$.38), corresponding to the typical full-width of the
image PSF for achieving the best S/N in creating the 1-dimensional (1D) spectra. For some cases small shifts of the aperture position on 
the sky were required. The 1D spectra from different orbits of an object have been averaged to obtain the final spectrum.
The calibration uncertainties of {\sl AKARI}/IRC spectra become large when $\lambda_{\rm obs} > 4.8~\micron$. We arbitrarily
excluded the $\lambda_{\rm obs} < 2.55~{\rm \micron}$ and $\lambda_{\rm obs} > 4.85~{\rm \micron}$ edges in order to avoid bad S/N data.  
Resultant spectra in the rest-frame wavelenght ($\lambda_{\rm rest} = \lambda_{\rm obs}/(1+z)$) are shown in figure
\ref{fig:all-spectra}. The aperture size corresponds to $\sim 2$ kpc at the distance of $\sim 100$ Mpc and thus our spectra are collected 
from regions weighted towards the central bulge-sized region around the nucleus and the contributions of disks are relatively suppressed.

\subsection{Correction for Galactic Extinction}
Before proceeding further, we have corrected our 2.5--5 $\micron$ spectra for Galactic extinctions as follows. Galactic extinctions in the
$K$-band at 2.2 $\micron$ (A$_{\rm K}$) were taken from the NED ExtraGalactic Catalog. Extinction values agree
with \citet{schlegel98} infrared-based dust map from the COBE/DIRBE and IRAS/ISSA which assumes a \citet{cardelli89} extinction law.
\citet{nishiyama06} determined the ratios of total to selective extinction in the IRSF/SIRIUS near-infrared (J,H,K$_{\rm S}$)
and  established that the extinction in the 2-3 $\micron$ wavelength range is well fitted by a power-law with a steep decrease
A$_{\lambda}\propto{\lambda^{-2}}$ toward the Galactic centre. We have made a small correction from A$_{\rm K}$ to A$_{\rm K_S}$
($\lambda_{\rm eff}=2.14~\micron$) using this relation. Then, we apply the adopted extinction law and the relation A$_{\lambda}/$A$_{\rm
K_S}$
\citep{roman07,nishiyama09} in order to do the flux correction considering the proper line-of-sight Galactic extinctions to the studied
AGNs across the whole 2.5--5 $\micron$ range. 

\subsection{The PAH Line Strength Measurements}

The software package MINUIT \citep{james75} has been used to obtain the fitted parameter values and errors
for the following analysis. By assuming a single Gaussian component for the 3.3 $\micron$ PAH emission feature we determined the peak, the
central wavelength and the dispersion $\sigma$ of the line profile based on the ${\chi}^2$ minimization over a local continuum. The
line flux is the integration over the Gaussian profile.

We have modeled the continuum in the rest-frame wavelength range between 3.15 and 3.35 $\micron$ with a power-law, if no notable feature
exists near 3.3 $\micron$. In some cases, there are nearby features such as the 3.1 $\micron$ H$_2$O ice covered dust and the PAH 3.4
$\micron$ sub-peak. In these cases,  we have included these features in the fitting process. In all cases with apparent PAH 3.3 $\micron$
emission feature, we see the 3.1 $\micron$ absorption. In some cases, we see the PAH 3.4 $\micron$ sub-peak. In order to determine the 3.3
$\micron$ feature parameters, we fit the rest-frame $2.75<\lambda {\rm [\micron]}<3.85$ spectrum [$f_{\rm rest}(\lambda)$] with the form:

\begin{eqnarray}
f_{\rm rest}(\lambda)&=& A_{\rm PL}\lambda^{-\Gamma}\,e^{-\tau_{3.1}{\;\mathrm{gauss}}(\lambda-\lambda_{3.1},\sigma_{3.1})}\nonumber \\
                              &+& f_{3.3}\;{\mathrm{gauss}}(\lambda-\lambda_{3.3},\sigma_{3.3})      \nonumber \\                           
                              &+& f_{3.4}\;{\mathrm{gauss}}(\lambda-\lambda_{3.4},\sigma_{3.4}).
\label{eq:fit}
\end{eqnarray}
The fitting parameters are $A_{\rm PL}$, $\Gamma$, $\tau_{\rm X}$,$f_{\rm X}$,$\lambda_{\rm X}$ and $\sigma_{X}$, where 
the subscript X (3.1,3.3 or 3.4) represents the nominal wavelength (in $\micron$) of the feature.
The first term represents the underlying power-law continuum with normalization $A_{\rm PL}$ and index $\Gamma$, multiplied
by the 3.1 $\micron$ ice covered dust absorption feature with an effective optical depth $\tau_{\rm 3.1~\micron}$. The second
and third terms represent the PAH 3.3 $\micron$ emission feature and the 3.4 $\micron$ sub-peak respectively with
line fluxes $f_{\rm X}$ and width $\sigma_{\rm X}$. The function 
$\mathrm{gauss}(\lambda,\sigma)=1/(\sqrt{2\pi}\sigma)\exp(-{\lambda^2}/{2\sigma^2})$ 
is the Gaussian function normalized to unity.  The central wavelengths are allowed to vary slightly near the nominal 
wavelength of each feature during the fit. The third term (the 3.4 $\micron$ PAH emission sub-peak) is included in
the fit if the sub-peak is clearly visible. 

From the subsample of detected 3.3 $\micron$ PAH emission features Gaussian $\sigma_{3.3}$ parameter was found to range 
from 0.025 $\micron$ to 0.04 $\micron$, with an average value of 0.030 $\micron$. For those objects for which  the PAH 3.3 $\micron$ is not
visible or only marginally visible,  we fixed the parameters $\lambda_{3.3}$ and $\sigma_{3.3}$ to $3.28$ and $0.030~\micron$ respectively
and investigated the variation of $\chi^2$ as a function of $f_{3.3}\geq 0$.  If the minimum $\chi^2$ (best-fit case) is smaller than  
2.7 below the $\chi^2$ value at $f_{3.3}=0$, we consider the line detected and report the best-fit $f_{3.3}$,  
otherwise, we consider it a non-detection  and report the 90\% upper limit to $f_{3.3}$  corresponding  to 
$\Delta \chi^2=2.7$ from the best-fit value.   

\begin{figure}
  \begin{center}
    \FigureFile(60mm,60mm){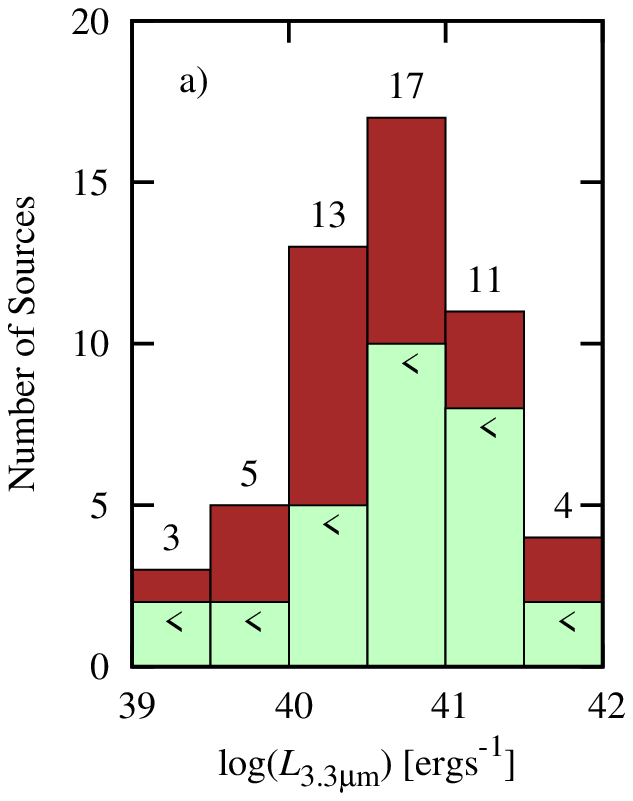}
    \FigureFile(60mm,60mm){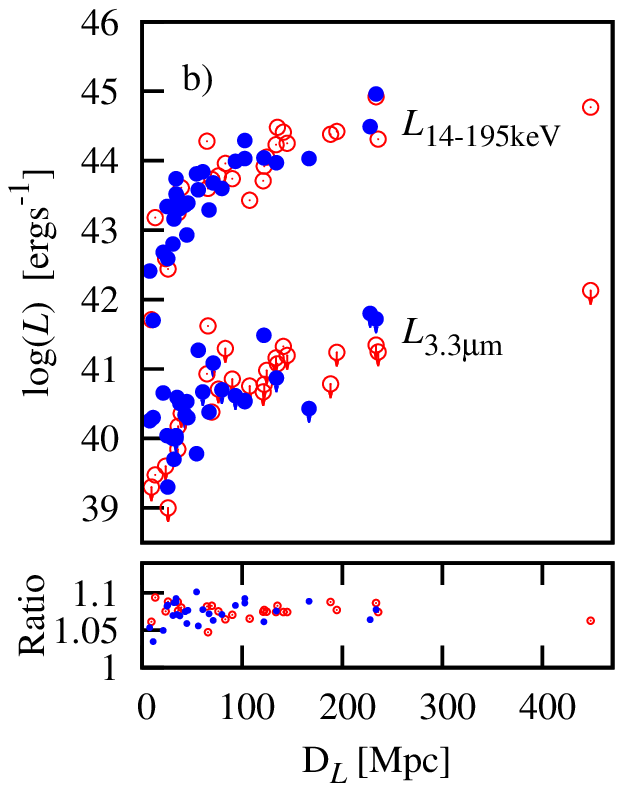}
  \end{center}
  \caption{(\emph{a}) Distribution of $\log(L_{\rm 3.3{\micron}})$ and (\emph{b}) luminosity-distance relationship of the
entire sample: Logarithm of the luminosity of the ${3.3{~\micron}}$ PAH emission in units of $\rm erg\,s^{-1}$(\emph{down}) and logarithm
of the hard-X ray luminosity in the 14 -- 195 keV band in units of $\rm erg\,s^{-1}$ (\emph{up}). The brown part shows actual detections
and the pale green part (with ``$<$'' symbols) show upper limits. Open red circles are optically classfied Seyfert 1 and solid blue circles
are Syeyfert 2s. Small downward arrows are upper-limits. (Color figures are available on the electronic version only.)}
\label{fig:lpah-distribution}
\end{figure}

All fitted fluxes and luminosities of the 3.3 $\micron$ PAH emission lines of the AGNs from our X-ray AGN selected sample are 
summarized in table \ref{tab:sample-fluxes}. For the PAH fluxes, 1$\sigma$ errors are reported for detections and 
the 90\% upper limits are reported for non-detections. We have converted our 3.3 $\micron$ flux to 
the line luminosity. The histogram of $\log(L_{3.3\micron})$ is shown in figure \ref{fig:lpah-distribution}(\emph{a}),
where upper limits are indicated. The values of $\log(L_{3.3\micron})$ and $\log(L_{\rm 14-195keV})$ 
are plotted as a function of luminosity distance in figure \ref{fig:lpah-distribution}(\emph{b}). 

\section{The Regression Analysis and Statistical Tests}\label{sec:regression-analysis}

The analysis of IRC spectra of our sample have both detections and non-detections of the 
PAH 3.3 $\micron$ feature. In this case, usual statistical techniques are no longer applicable. To study data containing both detections
and non-detections, we apply a series of survival analysis methods to the data using the ASURV package 
(\cite{feigelson85,isobe86,lavalley92}) to account for upper-limits (left censorship) of the 3.3 ${\micron}$ line luminosity. In table 
\ref{tab:regressions} we explored the correlation between   as well as the
correlation between the luminosities normalized by the black hole mass  ($M_{\rm {BH}}$). Masses presented in this work were
collected from the literature where mass is derived from the 2MASS $K$-band stellar magnitudes (\cite{mushotzky08,vasudevan09,winter09}).

Because of the presence of a scaling relation between the $M_{\rm {BH}}$ and 
the stellar mass of the bulge, the variable $L_{\rm 3.3{\micron}}/M_{\rm {BH}}$ can be considered a proxy for specific star formation rate
(SSFR). The variable $L_{\rm 14-195keV}/M_{\rm {BH}}$ is a proxy to Eddington ratio ($\lambda_{\rm edd}$). To test the difference of star
formation activities between different  types  of  AGNs  the  following correlation  analysis have  been  made.  First, based on
\citet{tueller08} optical classification, we divide our sample into two sub samples of optical type 1 AGNs and optical type 2
AGNs. Second, we repeated the analysis but in a column density classification scheme. We called X-ray type 1 AGNs the objects
with $N_{\rm H}\leq 10^{22}~{\rm cm^{-2}}$ and X-ray type 2 those objects with $N_{\rm H}>10^{22}~{\rm cm^{-2}}$.

In order to explore a posible correlation between the two variables the Cox regression method (where only the dependent variable have censored data) was employed. 
The parametric E-M (estimate and maximize) algorithm was used to determine the slope coefficients in a linear
regression model. This method is a general approach to the problem of finding maximum likelihood estimates for censored data sets \citep{isobe86} assuming a 
normal distribution of residuals. If censored data are not present this method yields the usual least-square results. 

To test the hypothesis that $L_{\rm 3.3{\micron}}/M_{\rm {BH}}$ for Seyfert 1 and Seyfert 2 objects have the same
distribution, the Gehan's extension of the Wilcoxon test, logrank test and Peto-Peto tests were used. We report the survival analysis
probabilities, P, from the mentioned tests in table \ref{tab:twost-results}. It shows that the probability that the distribution
of $\rm {3.3{~\micron}}$ PAH luminosities of Seyfert 1 and Seyfert 2 objects from our sample to be the same. A P value $\leq$ 0.05 means
that the two-subsamples differ at a statistically significant level, otherwise they are consistent with belonging to the same parent
population \citep{lamassa12}. Similar analysis was carried out for non-$M_{\rm BH}$ normalized data (see table \ref{tab:twost-results}). The
Kaplan-Meier estimator was used to obtain mean values for each sub-sample with the TWOST application under ASURV.
These tests are made for sub-samples divided by $L_{\rm 14-195 keV}/M_{\rm {BH}}$ and $L_{\rm 14-195 keV}$  for the 
tests for $L_{\rm3.3{\micron}}/M_{\rm {BH}}$  and $L_{\rm 3.3{\micron}}$ respectively. 

\chg{For the type 1 vs type 2 comparisons of $L_{\rm 3.3{\micron}}$ of the low $L_{\rm 14-195 keV}$ ($\rm{Low} -\it {L}_{\rm X}$) and high
$L_{\rm 14-195keV}$ ($\rm{High} -\it {L}_{\rm X}$) samples, we further verify the statistical significance of the comparisons using the
Bootstrap resampling method. We generate $N_{\rm boot}$ bootstrapped samples from each of the high and low $L_{\rm 14-195 keV}$ samples.
Each such bootstrapped sample contains the same number of objects ($n_{\rm obj}$) as the original sample and each object in the 
bootstrapped sample is a random selection from the original sample, in which an object in the original may be selected 
in duplicate.  The distribution of a statistical measure (e.g. mean value) from the $N_{\rm boot}$ redrawn samples is a good approximation
of that from samples (each with a size of $n_{\rm obj}$) randomly drawn from the underlying population. For each redrawn sample, we run the
TWOST application, which gives the mean $\langle \log L_{\rm 3.3\micron}\rangle$ for each of the type 1 and type 2 AGNs. Since our interest
is to see whether there is any systematic difference between type 1 and type 2 AGNs, we make a histogram of the difference  
$\langle \log L_{\rm 3.3\micron}\rangle _{\rm Sy1}-\langle \log L_{\rm 3.3\micron}\rangle _{\rm Sy2}$ from the $N_{\rm boot}=600$ 
bootstrapped samples to verify the significance of the difference.}

Since one of the major advantages of our sample is to have X-ray based $N_{\rm H}$ measurements for all AGNs, 
we can further explore the correlation using the $N_{\rm H}$ values rather the type 1/type 2 dichotomy. Thus we also 
investigate the correlation of $L_{\rm 3.3{\micron}}/M_{\rm {BH}}$ and $L_{\rm 3.3{\micron}}$ with $N_{\rm H}$. These tests should measure
the ``type'' or ``absorption'' dependence of SFR/SSFR without making somewhat arbitrary decisions about  type 1/type 2
borders.

\section{Results} \label{sec:results}

The results of our series of linear regression analyses using the E-M method are expressed
through the generic expression $ \log(A_i) =  a_i\;\;  \{ \log(B_i)-c_i \} \;+\; b_i$, where $A_i$ is the independent variable, $B_i$ is the
dependent variable, $a_i$ is the slope of the curve, $b_i$ is the abscissa intersection point, and $c_i$ is the average value of the
corresponding independent variable of the given relationship. The origin point of the distribution has been shifted to the average value of
the independent variable in order to minimize the artificial correlation of errors of the $a_i$ and $b_i$ parameters. This is needed because
ASURV does not provide the covariance matrix of parameter errors.  We have studied the dependencies between the luminosity of the PAH at 
$\lambda_{\rm rest}=$ 3.3 $\micron$ ($L_{\rm 3.3{\micron}}$) emission line to the X-ray luminosity in the 14--195 keV band ($L_{\rm
14-195keV}$; also refered to as $L_{\rm X}$):
\begin{equation} 
 \log(L_{\rm 3.3{\micron}}) =  a_{0}\;\;  \{ \log(L_{\rm 14-195keV})-c_{0}\} \;+\; b_{0}
 \label{ec:lum-no-normalized}
\end{equation}

Likewise, we express the relationship between the black-hole mass normalized luminosities:
\begin{equation}  
 \log(L_{\rm 3.3{\micron}}/M_{\rm {BH}}) =  a_{1}\;\;  \{ \log(L_{\rm 14-195keV}/M_{\rm {BH}})-c_{1}\} \;+\; b_{1}.
\label{ec:lum-normalized}
\end{equation}

A similar procedure was performed to explore a possible relationship between $N_{\rm H}$ and 
$L_{\rm 3.3{\micron}}$ ($L_{\rm 3.3{\micron}}/M_{\rm {BH}}$):

\begin{eqnarray}
\log(L_{\rm 3.3{\micron}})  =   a_{2}\;\;  \{ \log(N_{H})-  c_{2}\}  \;+\; b_{2}.            \label{ec:nh-corr1} \\
\log(L_{\rm 3.3{\micron}}/M_{\rm {BH}}) =   a_{3}\;\;  \{ \log(N_{H})-  c_{3}\}  \;+\; b_{3}. \label{ec:nh-corr2}
\end{eqnarray} 

\begin{figure}
  \begin{center}
    \FigureFile(80mm,80mm){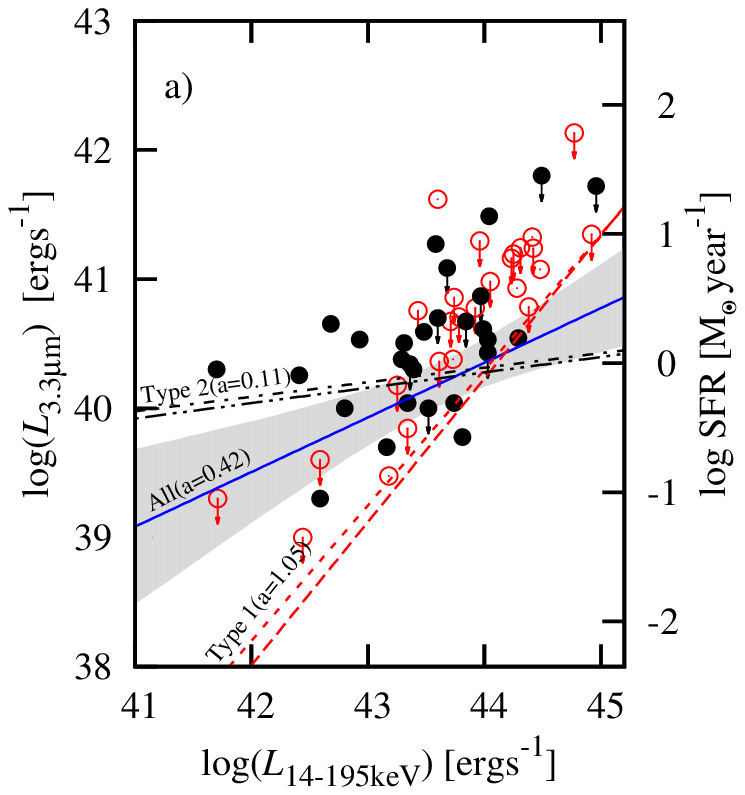}
    \FigureFile(80mm,80mm){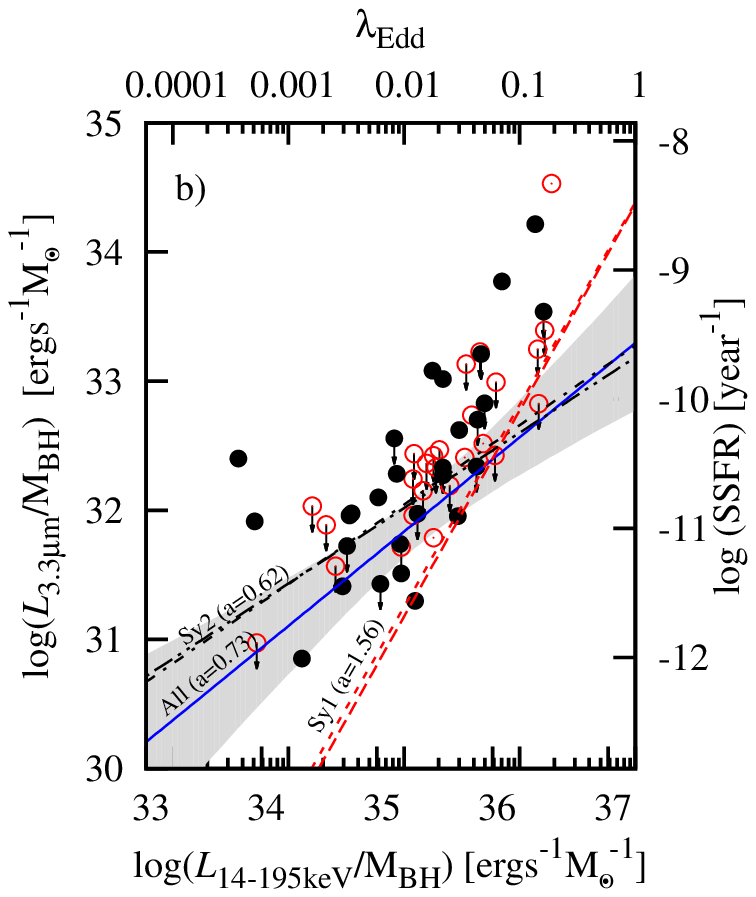}
  \end{center}
 \caption{(\emph{a}) $\log(L_{\rm 14-195keV})$ versus $\log(L_{\rm 3.3{\micron}})$ and  (\emph{b}) $\log(L_{\rm 14-195keV}/M_{\rm {BH}})$
versus $\log(L_{\rm 3.3{\micron}}/M_{\rm {BH}})$ relations.  On the right vertical axis of  panel (\emph{a}), approximate
star-formation rate corresponding to  $\log(L_{\rm 3.3{\micron}})$ are shown. Also on the upper horizontal and right vertical axes of
panel (\emph{b}), approximate Eddington ratios, $\lambda_{\rm Edd}$, corresponding to the $\log(L_{\rm 14-195 keV}/M_{\rm {BH}})$ values and
the specific star formation rate (SSFR) corresponding to $\log(L_{\rm 3.3\micron}/M_{\rm {BH}})$ are shown respectively. Arrows are for
upper-limits. In each figure, the best-fit linear regression for all-AGN sample is shown in solid blue lines, while the error region is
shown in gray shades.  Open red circles are used for optical type 1 objects and filled black circles for optical type 2 objects. Red dashed
and long-dashed lines corresponds to the regression lines for optical and X-ray type 1 objects, respectively. While the black dot-dashed and
dot-dot-dashed lines corresponds to the regression lines for optical and X-ray type 2 AGNs, respectively. 
(Color figures are available on the electronic version only.)}
\label{fig:correlations}
\end{figure}

The regressions have been made for all AGNs in our sample as well as for $L_{\rm 14-195 keV}$ and 
$L_{\rm 14-195 keV}/M_{\rm {BH}}$-divided sub-samples for equation \ref{ec:nh-corr1} and equation \ref{ec:nh-corr2} 
respectively.  The best-fit values and 1$\sigma$ errors for each equation coefficient are given by the 
ASURV package and summarized in table \ref{tab:regressions} and the scatter diagram with the best-fit 
lines are shown in figure \ref{fig:correlations}.

The average values of the independent variables used under this study are $c_0 = \langle\log L_{\rm 14-195keV}{\rm [{\rm
erg\,s^{-1}}]}\rangle=43.64$, $c_1 =
\langle\log L_{\rm 14-195keV}/M_{\rm {BH}} {\rm [{\rm erg\,s^{-1}}{M_{\odot}}^{-1}]}\rangle= 35.42$ and $c_2 = c_3 = \langle\log
(N_{\rm H}){\rm [cm^{-2}]}\rangle= 22.23$.

Based on our data analysis we do no find large discrepancy between the X-ray and optical classifications. Almost half of the sources, 26/54
(48\%), have optical classifications of Sy 1-1.5. The mean X-ray column density for these objects corresponds to a low column
density object (un-absorbed) with a $\log N_{H} = 20.83$. The 1.6-2.0 optically classified sources (28/54; 52\%) have as expected a higher
column density, $\log N_{H} = 23.35$. When we use the X-ray criteria the proportion is similar: 24 X-ray type 1 sources (44\%) and 30 X-ray
type 2 sources (56\%).

We applied the generalized Cox's proportional hazard model to compute the correlation probabilities along with the E-M algorithm which
calculates the linear regression coefficients. The results of the regressions are summarized in table \ref{tab:regressions}.

As shown in table \ref{tab:regressions},  the probabilities that a correlation is not present for the
$\log(L_{\rm 14-195keV})$ versus $\log(L_{3.3{\micron}})$ relationship is 0.01 and for $\log(L_{\rm 14-195keV}/M_{\rm {BH}})$ versus 
$\log(L_{\rm 3.3{\micron}}/M_{\rm {BH}})$ is 0.002, implying that a correlation is present through the whole sample. 
We divided the sample according to the optical classification of the sources. The probability that a correlation in not present for
the optical Seyfert 1 objects is 0.02 for the $\log(L_{\rm 14-195keV})$-$\log(L_{\rm 3.3{\micron}})$ relationship and 0.005 for the $M_{\rm
BH}$ normalized case, indicating significant correlations. For the optical Seyfert 2's, the probability that there is no correlation between
$L_{\rm 14-195keV}$ and $L_{\rm 3.3{\micron}}$ is 0.65, while the same probability is 0.1 between $\log(L_{\rm 14-195keV}/M_{\rm BH})$ and
$\log(L_{\rm 3.3{\micron}}/M_{\rm BH})$. 
Thus no significant correlation has been found between the AGN power and star-formation rate in Seyfert 2 galaxies.
The correlation is marginal in the normlized case for Seyfert 2's. 

The sample has been also subdivided according to a X-ray column density classification scheme instead of the optical classification. We call
the sources with $N_{\rm H}\leq 10^{22}~{\rm cm^{-2}}$ ``X-ray type 1 AGNs'' and those with 
$N_{\rm H}>10^{22}~{\rm cm^{-2}}$ ``X-ray type 2 AGNs''. The results in table \ref{tab:regressions} show that the difference between
correlations in optical and X-ray AGN type division schemes are different by only about 4\% from each other. Figure \ref{fig:correlations}
shows the scatter diagrams between $\log(L_{\rm 14-195keV})$ and $\log(L_{\rm 3.3{\micron}})$ as well as between  $\log(L_{\rm
14-195keV}/M_{\rm BH})$ and $\log(L_{\rm 3.3{\micron}}/M_{\rm BH})$. The best-fit regressions for the all-AGN sample as well as type-divided
samples are shown. The error range of the regression line corresponding to $\Delta \chi^2<2.3$ (68\% confidence for the two interesting
parameters) is also shown as a shaded area in each panel for the all-AGN sample. In both figures, the regression line of type 1 AGNs  shows
a steeper slope than that of type 2 AGNs.  The tendency is common for optically-divided types and X-ray divided types. The differences of
the slopes between type 1 and type 2 regression curves are at the 2-3$\sigma$ levels. 

To further verify this tendency, we have made further statistical
tests. We divided the sample into high and low $L_{\rm 14-195 keV}$ (or $L_{\rm 14-195 keV}/M_{\rm {BH}}$) and compared the mean 
$\log L_{\rm 3.3{\micron}}$ (or $L_{\rm 3.3 \micron}/M_{\rm {BH}}$) values of the type 1 and type 2 AGNs 
(see table \ref{tab:twost-results}) using a number of two-sample tests available in ASURV.   
We used the Gehan's Generalized Wilcoxon test, logrank test and Peto \& Peto Generalized Wilcoxon Test 
to determine the probability that the distributions of (S)SFR proxy among the type 1 and type 2 sub-samples 
are drawn from the same parent population separately for high and low $L_{\rm 14-195 keV}$ (or $L_{\rm 14-195 keV}/M_{\rm {BH}}$) 
regimes. The only statistically significant difference between the type 1 and type 2 samples in these two-sample tests
are in the $\langle \log(L_{\rm 3.3{\micron}}) \rangle$ values of low X-ray liminosity sample. The difference is marginal in the
$M_{\rm BH}$ normalized case.

\begin{figure}
  \begin{center} 
    \FigureFile(70mm,70mm){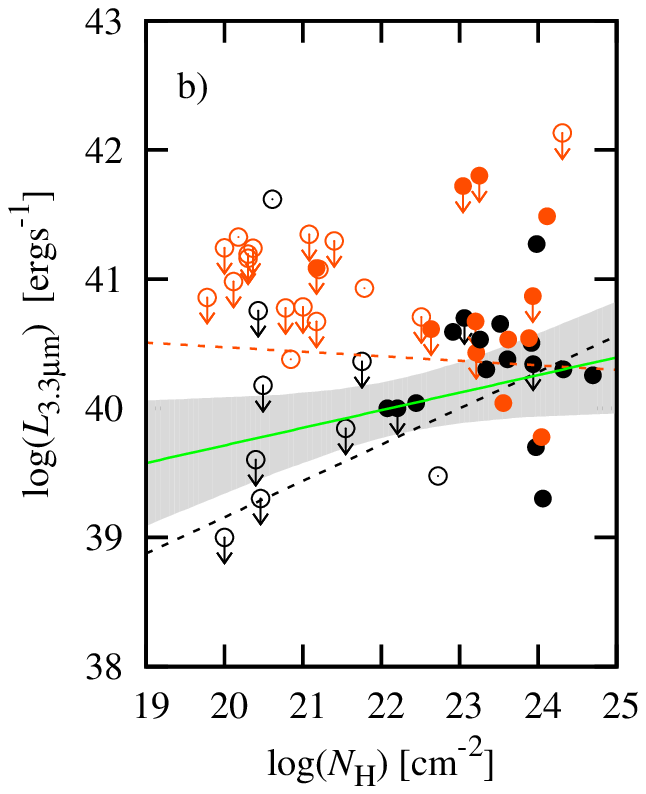}
    \FigureFile(70mm,70mm){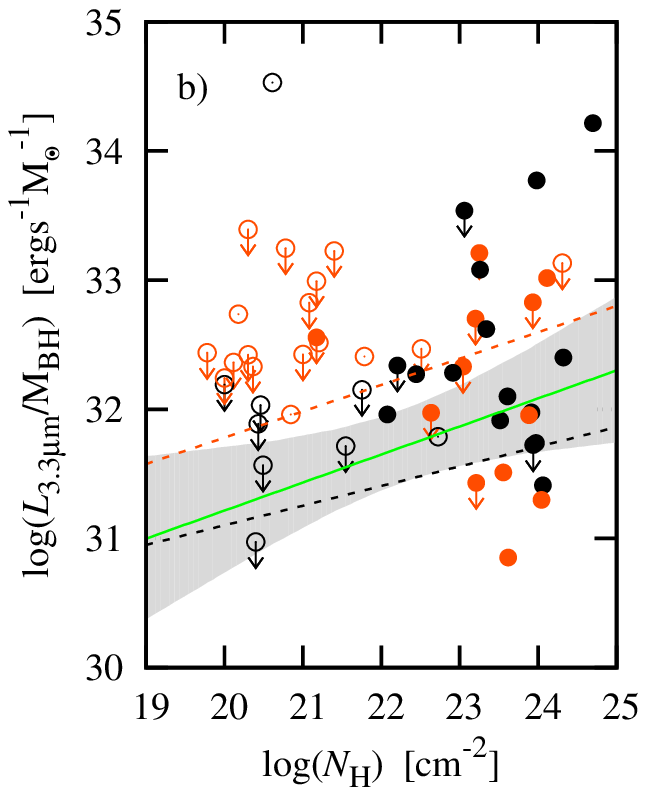}
  \end{center}
  \caption{ (\emph{a}) $\log(L_{\rm 3.3{\micron}})$ versus $N_{\rm H}$ relationship. Orange (paler) symbols have been used for AGNs with
$\log(L_{\rm 14-195keV})   > 43.64$ and black for $\log(L_{\rm 14-195keV}) \le 43.64$, orange and black dashed lines are the regression
fits for the subsamples, respectively. (\emph{b}) $\log(L_{\rm  3.3{\micron}}/M_{\rm {BH}})$ versus $N_{\rm H}$ relationship    Orange
symbols have been used for AGN with $\log(L_{\rm 3.3\micron}/M_{\rm BH})> 35.43$ and black for $\log(L_{\rm 3.3\micron}/M_{\rm BH}) \le
35.43$ , orange and black dashed lines are the regression fits for the subsamples, respectively. Open circles are for X-ray type 1 objects
and filled circles for X-ray type 2 objects.   Arrows are for upper-limits. Solid green lines are the linear fits for the whole sample in
each case.  For numerical details see table \ref{tab:regressions}. (Color figures are available on the electronic version only.)}
\label{fig:nh-classified}
\end{figure}

The basic results of the regressions involving $N_{\rm H}$ (see equations (\ref{ec:nh-corr1}) and (\ref{ec:nh-corr2})) are as follows. 
We do not find significant correlations between  $\log(N_{\rm H})$ and $\log L_{\rm 3.3{\micron}}$ for the all-AGN sample
(see figure \ref{fig:nh-classified}(\emph{a})). However, if we divide the sample in two X-ray luminosity bins, a positive correlation has
been observed in only low luminosity AGNs.

No significant correlation has been found between $\log({N_{\rm H}})$ and $\log(L_{\rm 3.3{\micron}}/M_{\rm {BH}})$ 
relationship in any of the all, high $\log(L_{\rm 3.3{\micron}}/M_{\rm {BH}})$ and low $\log(L_{\rm 3.3{\micron}}/M_{\rm {BH}})$ samples  
(see figure \ref{fig:nh-classified}(\emph{b}).

On the right vertical axis of figure \ref{fig:correlations}(\emph{a}) the SFR, which is estimated by using the   $L_{\rm FIR}-L_{\rm
3.3 \micron}$ relation by \citet{mouri90} and the $L_{\rm FIR}-$SFR relation by \citet{kennicutt98} using:

\begin{equation}
 \log (SFR)[\rm M_{\odot}year^{-1}] =  \log (L_{\rm 3.3{\micron}}{[\rm erg\,s^{-1}]})\;-\;40.34
\label{ec:SFR}
\end{equation}

On the upper horizontal axis of figure \ref{fig:correlations}(\emph{b}), approximate Eddington ratios $\lambda_{\rm Edd}\equiv L_{\rm
bol}/L_{\rm Edd}$ (see equation \ref{ec:lambda_edd}), where $L_{\rm bol}$ is the bolometric luminosity of the AGN and $L_{\rm
Edd}=1.26\times 10^{38}(M_{\rm BH}/M_{\rm \odot}){\rm erg\,s^{-1}}$ is the Eddington luminosity, corresponding to $L_{\rm 14-195 keV}/M_{\rm
BH}$ values are indicated. The conversion has been made as follows.
First, we convert from the 14--195 keV to unabsorbed 2-10 keV luminosity using an effective photon index of $\Gamma=1.85$,
which implies $L_{\rm 2-10keV}/L_{\rm 14-195 keV}=0.41$. This is based on \citet{ueda11}, where average effective photon index
between these two bands range from $\Gamma\approx 1.7$ at the low luminosity end to $\Gamma \approx 2.0$ in the low luminosity end.
 For the bolometric correction from 2-10 keV, we use $L_{\rm bol}/L_{2-10 keV}=14$, from \citet{lusso12} for 
$\log(L_{\rm 2-10 keV}) \approx 43.2$ type 1 AGNs, which is the average 2-10 keV unabsorbed luminosity of AGNs in our sample implied from
the mean $\langle (\log L_{\rm 14-195 keV}) \rangle$. 
\begin{equation}
 \lambda_{\rm Edd} = L_{\rm Bol}/L_{\rm Edd} \sim 5\times 10^{-38}(L_{\rm 14-195 keV}{\rm [erg\;s^{-1}]}/M_{\rm BH}[\rm M_{\odot}])
\label{ec:lambda_edd}
\end{equation}

The rough SSFR scale on the right axis of figure \ref{fig:correlations}(\emph{b}) is determined based  
on the combination of equation \ref{ec:SFR}, the $M_{BH}-L_{K(\rm stellar)}$ relation from \citet{mushotzky08} and 
the stellar mass to $K$-band luminosity ratio, $M_{\rm stellar}/L_{\rm K(stellar)} \sim 0.8$ (in solar units) \citep{brinchmann00}. 
The $M_{\rm BH}$ dependence of the ratio $M_{BH}/L_{K(\rm stellar)}$ is neglected and is
evaluated at $\log(M_{\rm BH})=8.27$, which is the mean value for our sample.

\begin{equation}
 \log (SSFR)[\rm year^{-1}] =\log (L_{\rm 3.3{\micron}}[\rm erg\,s^{-1}])-\log (M_{BH}[M_{\rm \odot}])\;-\;42.87
\label{ec:SSFR}
\end{equation}

Since there is significant scatter and luminosity/mass dependence in the conversions involved, these relations are only accurate 
to an order of magnitude.

\begin{figure}
  \begin{center} 
    \FigureFile(90mm,90mm){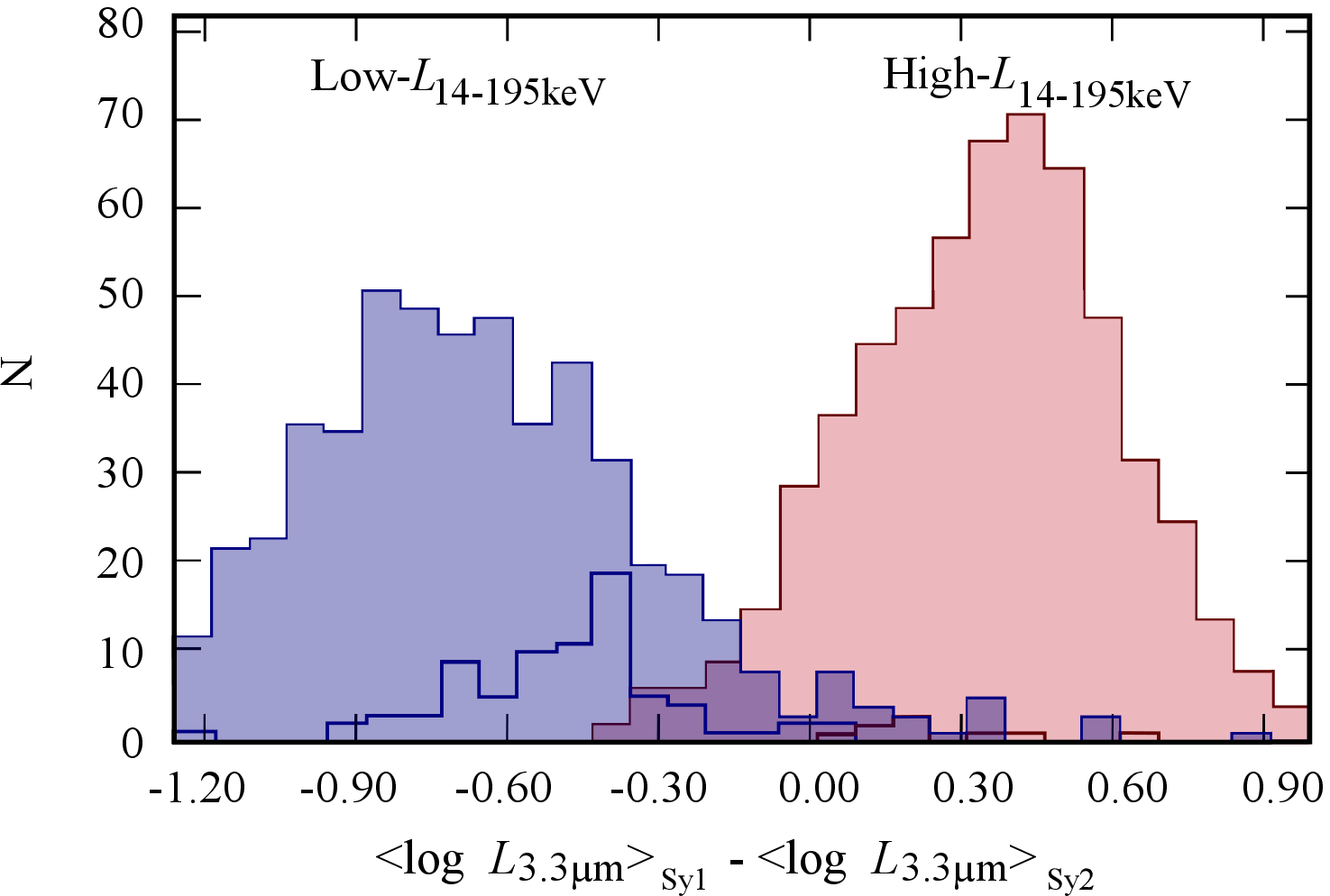}
  \end{center}
  \caption{The bootstrap histograms of $\Delta_{\rm 12}=\langle \log L_{\rm 3.3\micron}\rangle_{\rm Sy1}-\langle \log L_{\rm 3.3\mu
m}\rangle_{\rm Sy2}$ are shown for the high $L_{\rm 14-195 keV}$ (blue/darker histogram) and low $L_{\rm 14-195 keV}$ (red/paler histogram)
samples. The histograms below thick solid lines, which are labeled as a number of "$\textless$"'s show the cases where the TWOST routine
fails to give $\langle \log L_{\rm 3.3\micron}\rangle_{\rm Sy1}$ due to too many upper limits in the corresponding redrawn sample,
in which the upper limits are used for the mean calculations. (Color
figures are available on the electronic version only.) }
\label{fig:bootstrap}
\end{figure}

\begin{figure}
  \begin{center} 
    \FigureFile(70mm,70mm){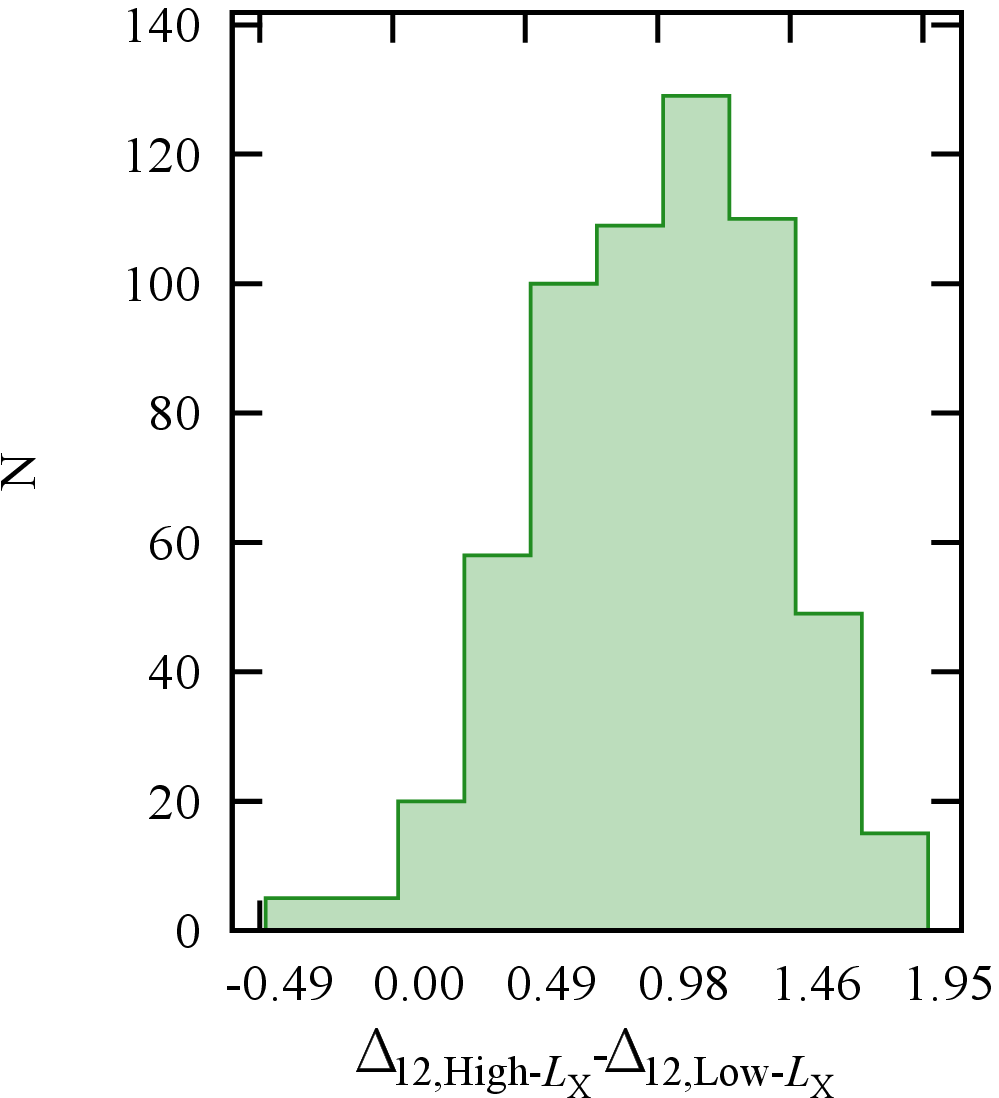}
  \end{center}
  \caption{Histogram of the difference ($\Delta_{12,\rm{High} -\it {L}_{\rm X}}-\Delta_{12,\rm{Low} -\it {L}_{\rm X}}$) for 600 randomly
selected high $L_{\rm 14-195 keV}$ and low $L_{\rm 14-195 keV}$ pairs from re-drawn samples, respectively. The distribution of
($\Delta_{12,\rm{High} -\it {L}_{\rm X}}-\Delta_{12,\rm{Low} -\it {L}_{\rm X}}$) shows that the probability that it becomes less than zero
by chance is only 0.75\%.(Color figures are
available on the electronic version only.)}
\label{fig:intersected}
\end{figure}

\chg{
The most significant result of our tests is the excess of $L_{\rm 3.3\micron}$ of type 2/absorbed AGNs with respect
to that of type 1/unabsorbed AGNs at low $L_{\rm 14-195 keV}$($\rm{Low} -\it {L}_{\rm X}$). However, this excess is not observed at high
$L_{\rm 14-195 keV}$ ($\rm{High} -\it {L}_{\rm X}$). These results are worth scrutinizing and therefore we made bootstrap resampling to the
each of the high and low  $L_{\rm 14-195 keV}$ samples as described in section \ref{sec:regression-analysis}. The bootstrap histograms of 
$\Delta_{\rm 12}\equiv \langle \log L_{\rm 3.3\micron}\rangle _{\rm Sy1} - \langle \log L_{\rm 3.3\micron}\rangle _{\rm Sy2}$ for 600
redrawn samples for each of the high and low $\log L_{\rm 14-195 keV}$ sub-samples are shown in figure \ref{fig:bootstrap}. 
In some redrawn samples where there are too many upper limits, the TWOST routine cannot determine the mean 
$\langle \log(L_{\rm 3.3\micron}) \rangle _{\rm Sy1}$ value and instead gives an NaN (not a number). There are 9/600 and 78/600 
such for the high and low $L_{\rm 14-195 keV}$ samples respectively. In these cases, we use the upper limit values 
to calculate the mean. The histograms of these cases are also overplotted in figure \ref{fig:bootstrap} under thick 
lines and indicated by symbols '$<<<<<<$'. 

The bootstrap histogram shows that only 29 out of 600 bootstraps (5\%)
show $\langle \log L_{\rm 3.3\micron}\rangle _{\rm Sy1} - \langle \log L_{\rm 3.3\micron}\rangle _{\rm Sy2}>0$ for the low 
$L_{\rm 14-195 keV}$ sample, verifying the conclusion of the TWOST tests. This percentage is an overestimate
considering that $\sim 4$ of the 29 $\Delta_{\rm 12}>0$ cases are upper limits. For the high $L_{\rm 14-195 keV}$ sample,
where the mean $\Delta_{\rm 12}$ is positive, 58 cases out of 600 bootstraps give $\Delta_{\rm 12}<0$. Thus the SFR in more enhanced in
type 1 than in type 2 sources in the high X-ray luminosity sample with only a marginal significance. 

One important question is whether there is any systematic difference of $\Delta_{12}$ values between high and low $L_{\rm 14-195keV}$
samples. In order to test whether the $\Delta_{\rm 12}$ is significantly different between the high and low $L_{\rm 14-195 keV}$ samples, we
calculated the difference ($\Delta_{12,\rm{High} -\it {L}_{\rm X}}-\Delta_{12,\rm{Low} -\it {L}_{\rm X}}$) for 600 randomly selected high
X-ray luminosity-low X-ray luminosity pairs from re-drawn samples (see figure \ref{fig:intersected}). The distribution of 
($\Delta_{12,\rm{High} -\it {L}_{\rm X}}-\Delta_{12,\rm{Low} -\it {L}_{\rm X}}$) shows that the probability that it becomes less than zero
by chance is only 0.75\%.

}

\section{Discussion/Future Plan} \label{sec:discussion}
Astronomical surveys are often affected by a selection effect derived from a preferential detection of intrinsically bright objects.  The
luminosities of detected objects found in a flux-limited survey presents a strong distance dependence. Lower luminosity objects at high
redshift tend to be censored due to the sensitivity limitations of the instruments. \citet{feigelson83} argued that if censored data is
properly treated through the use of survival analysis methods, one can remove the redshift dependence from the luminosity relation. 
Results from simulations presented in \citet{feigelson85} tend to support these assertions. Cox's test for correlation can remove the
selection effect and recover the latent relationship between the involved variables. Our regression with the E-M algorithm show significant
positive correlation
between $\log(L_{\rm 14-195keV}/M_{\rm {BH}})$ and $\log(L_{\rm 3.3{\micron}}/M_{\rm {BH}})$ as well as between $\log(L_{\rm 14-195keV})$
and
$\log(L_{\rm 3.3{\micron}})$, thus there seem to be real underlying correlation between the AGN and circum-nuclear AGN
activities. Another effect that might cause spurious correlations is an aperture effect, where more distant 
AGNs include more star-formation activities from off-circum-nuclear region such as disks. We estimate the degree of this effect
by  extracting spectra of our nearby AGNs with apertures that cover the entire galaxy. Typically the PAH luminosity increases
by a factor of two, where the increase due to finite point spread function of {\sl AKARI} is $\sim 30$\%, which is
estimated from the continuum at $\sim$ 3.3 $\micron$ of the QSO 3C 273. Since the correlations extends over $\sim 1.5$ orders 
of magnitude in both $\log(L_{\rm 3.3{\micron}})$ or $\log(L_{\rm 3.3{\micron}}/M_{\rm {BH}})$, this aperture effect does not alter our
correlation results significantly.      

\citet{woo12} investigated the connection between starburst and AGN activity by comparing the 3.3 $\micron$ PAH emission 
and AGN properties of a more distant sample ($z \sim 0.4$) of moderate-luminosity Seyfert 1s. The 3.3 $\micron$ feature was 
detected in 7 of 26 target galaxies. They found no strong correlation between the ${\rm 3.3{\micron}}$ global emission of 
PAH and AGN luminosity at 5100 ${\textup{\AA}}$. Their sample is enclosed within a fairly narrow range of luminosity and 
little information can be concluded from these
observations. However, by combining with data from literature and assuming a fixed scaling relationship between global emission of 3.3
$\micron$ PAH and nuclear 3.3 $\micron$ PAH emission (based on {\sl AKARI}/IRC observations of NGC 7469 performed by \citet{imanishi10} and
ground-based spectrograph with a narrow slit of 1$^{\prime\prime}$.6 by \citet{imanishi04}, respectively), they found a
correlation with the luminosity of the AGN on a wider liminosity span, suggesting that star formation and AGN activity could be closely
related in the nuclear region. Their adopted flux ratio between nuclear $L_{\rm 3.3{\micron}}$ and the global $L_{\rm 3.3{\micron}}$ at $z
\sim 0.4$ 
is 0.04. 

As found in \citet{oi10}, we see no clear difference in the $\langle \log(L_{3.3\micron}) \rangle$ between the two types of AGNs 
for our overall sample.  Neither do we find significant difference in the $\langle \log(L_{3.3\micron}/M_{\rm {BH}}) \rangle$ between 
type 1 and type 2 AGNs if AGNs in all luminosities are included. However, our regression analysis show that type 1 AGNs 
exhibit steeper slope in the scatter diagram of $\log(L_{\rm 3.3\micron})$ plotted as a function of $\log(L_{\rm 14-195 keV})$. 
The same trend has been found for the $\log(L_{\rm 3.3\micron}/M_{\rm {BH}})$ plotted 
as a function of $\log(L_{\rm 14-195 keV}/M_{\rm {BH}})$. We find that the mean $\log(L_{3.3\micron})$ value is significantly larger in the
type 2 AGNs than that of type 1 AGNs, if we limit the sample to lower X-ray luminosity AGNs ($\log(L_{\rm 14-149 keV}) \leq 43.64$) while
we find no ststistically significant difference for the higher luminosity AGNs ($\log(L_{\rm 14-149 keV})>43.64$). A similar trend
has been found in the comparison between $\log(L_{\rm 3.3\micron}/M_{\rm {BH}})$ and $\log(L_{\rm 14-195 keV}/M_{\rm {BH}})$ with a lower
statistical significance. We also find a positive correlation between $N_{\rm H}$ and $\log(L_{\rm 3.3\micron})$ for the low
X-ray luminosity sample only, while no significant correlations have been found between $N_{\rm H}$ and $\log(L_{\rm 3.3\micron}/M_{\rm
BH})$.  

In summary, our analysis found enhanced star-formation for low X-ray luminosity type 2 
Seyferts than type 1 Seyferts, while we find no significant difference in high X-ray luminosity AGNs. 
Thus the Seyfert type dependence of the SFR is luminosity dependent. As seen in Fig. \ref{fig:intersected}, the
X-ray luminosity dependence of the AGN type versus SFR relation is statistically robust.

 Although it is highly speculative yet, one may interpret this observation as follows. In the low luminosity AGNs, the difference 
between type 1 and type 2 AGNs may reflect an evolution sequence, where in the early stage of AGN activity, the kpc-scale circum-nuclear 
star formation, which feeds the central black hole still remains and therefore the AGNs are still surrounded by thick torus, which has 
a higher chance to be observed as type 2 AGNs. As the starburst fades away which may or may not be quenched by the AGN feedback, the scale 
height of the torus gets lower and they have more chance to be observed as type 1's. 
\chg{On the other hand, the situation in high luminosity AGNs might be different. It is well known that the type 2 or absorbed
AGN fraction (number density of those with $N_{\rm H}=10^{22-24}{\rm cm^{-1}}$ to that of $N_{\rm H}<10^{24}{\rm cm^{-1}}$)
among X-ray AGNs (excluding highly unexplored population of Compton-thick AGNs with $N_{\rm H}>10^{24}{\rm cm^{-1}}$) decreases
with X-ray luminosity (e.g. \cite{hasinger08,ueda14}). Thus, on average, the X-ray high luminosity AGNs are surrounded by
a thinner torus and it is not unreasonable to assume that the dispersion of the torus opening angles is smaller 
at high X-ray luminosities than at lower luminosities. Thus at high luminosities, the difference between type 1's and type 2's might be
mainly caused by the line-of-sight effect.}

An underlying assumption of the survival analysis is that the intrinsic scatter around the best-fit
line (regression) or the mean value (two sample tests) is gaussian, which is not guraranteed. Thus it is important to
confirm (or deny) our results with as few upper limits as possible.

In our future paper, we will extend our analysis to Spitzer IRS spectroscopy from the archive to utilize the PAH features at 6.2 $\micron$,
7.7 $\micron$ $\micron$, 11.3 $\micron$ and 17 $\micron$. By involving these PAH lines, we will be able to reduce the number of upper-limits
for more robust conclusions.

\section{Conclusions} \label{sec:conclusions}

 We investigate the 2.5--5 $\micron$ spectra of 54 bright nearby non-blazar AGNs from the 9-month {\sl Swift} BAT catalog using 
{\sl AKARI}/IRC. We investigate the relation between AGN type/absorption and star formation activities. From our present work, 
we conclude the following:
\begin{itemize}
\item We have detected ${\rm 3.3{~\micron}}$ PAH emission from 24 out of 54 flux limited sample of hard X-ray selected AGNs. 
\item Strong correlations have been found between $\log(L_{\rm 14-195keV})$ and $\log(L_{\rm 3.3{\micron}})$ as well
as between $\log(L_{14-195keV}/M_{\rm {BH}})$ and $\log(L_{\rm 3.3{\micron}}/M_{\rm {BH}})$ for both optical and X-ray classified type 1
AGNs.
\item We have found no statistical difference in the mean circum-nuclear SFR, traced by the PAH 3.3 $\micron$ emission, 
  between type 1 and type 2 AGNs for our overall sample.
\item If we limit ourselves to low luminosity AGNs, we have stronger nuclear starburst activity in type 2 AGNs than type 1 AGNs. 
  There is no significant difference in the star-formation activity betwen high luminosity type 1 and type 2 AGNs.
\item A similar trend has been found for the SSFR, between low and high Eddington ratio samples,
  although the statistical significance is lower.
\item Significant correlation have been found between $\log(N_{\rm H})$ and $\log(L_{\rm 3.3\micron})$ for the Low-$L_{\rm X}$ sample,
  while no significant correlations have been found for the high $L_{\rm X}$ sample.
  The significance of correlations between $\log(N_{\rm H})$ and $\log (L_{\rm 3.3\micron}/M_{\rm BH})$ in any sample are much weaker, if
any.
\item Our results suggest that the difference between type 1/type 2 in low luminosity AGNs may reflect an evolution
  sequence, where more obscuring material is available around low luminosity type 2 AGNs when the circum-nuclear star-formation
  is feeding the central engine. At high luminosities, the difference between the two types may be mainly from the orientation
  effect.  
\item Our analysis depends on the validity of the survaival analysis in the presence of upper limits of 
  the 3.3 $\micron$ luminosities. Our findings have to be confirmed with other measures of the star formation activity 
  that are not contaminated by the AGNs, such as other PAH features measured with {\sl Spitzer} IRS.
\end{itemize}

\bigskip
\bigskip
This work has been supported by CONACyT Grant 179662 and DGAPA-Universidad Nacional Aut\'onoma de M\'exico (UNAM) Grant PAPIIT 
IN104113. This work is also supported by the Grant-in-Aid for Scientific Research 23540273 (MI) and 26400228 (YU) from the
Ministry of Education, Culture, Sports, Science and Technology of Japan (MEXT). This research is based
on observations with {\sl AKARI}, a Japan Aerospace Exploration Agency (JAXA) project with the 
participation of ESA. The {\sl Swift}/BAT 9-month cataloge site is managed by the NASA Goddard Space Flight Center.

\newpage
\begin{longtable}{lccc}
  \caption{{\sl AKARI}/IRC Observation log for hard X-ray selected AGNs}\label{tab:obs-log}
  \hline           
{\sl Swift}/BAT Name & Counterpart Name & Observation ID& Observation Date\\

\endfirsthead
  \hline  
  \hline
\endhead
  \hline
\endfoot
  \hline
\endlastfoot
  \hline
   SWIFT J0048.8+3155 & NGC 262                  &  1122156-1			& 2010-01-15		\\
   SWIFT J0123.9-5846 & Fairall 9		 &  1340445-1,3			& 2008-12-01,02		\\
   SWIFT J0134.1-3625 & NGC 612                  &  1120076-1,2,4,5		& 2008-06-24,26		\\  
   SWIFT J0138.6-4001 & ESO 297-018              &  1120074-1,2,3,4,5 		& 2008-06-23		\\  
   SWIFT J0214.6-0049 & Mrk 590	                 &  1340446-1,2			& 2009-07-24		\\
   SWIFT J0238.2-5213 & ESO 198-024              &  1122056-1,2,3,4,5 		& 2009-12-25		\\  
   SWIFT J0319.7+4132 & NGC 1275                 &  1120056-1			& 2009-08-21		\\
   SWIFT J0426.2-5711 & 1H 0419-577	         &  1920103-1,2			& 2009-01-16		\\
   SWIFT J0433.0+0521 & 3C 120	                 &  1340447-1,2,3		& 2009-02-25		\\
   SWIFT J0516.2-0009 & Ark 120	                 &  1340448-1,2,3		& 2008-09-09		\\
   SWIFT J0519.5-3140 & ESO 362-G021	         &  1920114-1,2			& 2009-03-04		\\
   SWIFT J0554.8+4625 & MCG+08-11-011            &  1120063-1,2,3		& 2009-09-21		\\
   SWIFT J0601.9-8636 & ESO 005-G004             &  1120073-1,2,3,4,5		& 2008-09-19,20		\\
   SWIFT J0615.8+7101 & Mrk 3                    &  1120001-1,2,3		& 2008-09-23		\\
   SWIFT J0623.9-6058 & ESO 121-G028	         &  1122044-1,2,3,5		& 2009-10-18,24		\\
   SWIFT J0651.9+7426 & Mrk 6                    &  1120064-1,2,3		& 2008-09-27		\\
   SWIFT J0742.5+4948 & Mrk 79		         &  1340470-1,2,3		& 2008-10-10		\\
   SWIFT J0902.0+6007 & Mrk 18                   &  1122043-1,2,3,4,5		& 2009-10-20		\\
   SWIFT J0920.8-0805 & MCG-01-24-012	         &  1122045-1			& 2009-11-17		\\
   SWIFT J0925.0+5218 & Mrk 110	                 &  1340451-1,2,3		& 2009-04-25		\\
   SWIFT J0945.6-1420 & NGC 2992		 &  3750049-1,2,3		& 2009-11-25,26		\\
   SWIFT J0947.6-3057 & MCG-05-23-016            &  1122050-1,2,3		& 2009-12-03,05		\\
   SWIFT J0959.5-2248 & NGC 3081                 &  1120082-1,2,3,4,5		& 2009-06-02		\\
   SWIFT J1031.7-3451 & NGC 3281                 &  1120075-1,2			& 2009-06-17		\\
   SWIFT J1049.4+2258 & Mrk 417                  &  1120083-1,2			& 2009-05-26		\\
   SWIFT J1106.5+7234 & NGC 3516                 &  1122032-1,2,3,4		& 2009-10-25		\\
   SWIFT J1139.0-3743 & NGC 3783		 &  1340453-1,2,3		& 2008-07-03		\\
   SWIFT J1143.7+7942 & UGC 06728                &  1122054-1,2,4,5		& 2009-10-15		\\
   SWIFT J1203.0+4433 & NGC 4051		 &  1340473-1			& 2009-05-29		\\
   SWIFT J1206.2+5243 & NGC 4102                 &  1120232-1,1122090-1		& 2009-05-25,2009-11-27	\\
   SWIFT J1210.5+3924 & NGC 4151                 &  1122024-1,1340454-1,2,3	& 2008-06-03,2009-12-05	\\
   SWIFT J1225.8+1240 & NGC 4388                 &  1120080-1,2,3		& 2009-06-21		\\
   SWIFT J1238.9-2720 & ESO 506-G027             &  1120078-1,2,3,1120079-1,2	& 2009-01-10,2009-07-11	\\
   SWIFT J1239.6-0519 & NGC 4593		 &  1340475-1,2			& 2008-07-02,2009-01-01	\\
   SWIFT J1303.8+5345 & SBS 1301+540	         &  1122053-1,2,3,4,5		& 2009-12-04,05		\\
   SWIFT J1305.4-4928 & NGC 4945	  	 &  3180009-1			& 2007-01-27		\\
   SWIFT J1322.2-1641 & MCG-03-34-064            &  1120084-1,2,3,4,5		& 2008-07-17		\\
   SWIFT J1338.2+0433 & NGC 5252                 &  1120085-1,2,3,4,5		& 2009-07-12,13		\\
   SWIFT J1349.3-3018 & IC 4329A		 &  3750054-1,2,3		& 2010-01-26		\\
   SWIFT J1352.8+6917 & Mrk 279	                 &  1340458-1,2,3		& 2008-11-14,15,16	\\
   SWIFT J1413.2-0312 & NGC 5506                 &  1120068,1,2 		& 2008-07-24		\\
   SWIFT J1417.9+2507 & NGC 5548		 &  1340460-1,2,3		& 2008-07-13,14		\\
   SWIFT J1442.5-1715 & NGC 5728                 &  1120086-1,2,3		& 2009-08-05,06		\\
   SWIFT J1535.9+5751 & Mrk 290		         &  1340550-1,2,3		& 2008-06-29,2009-01-01	\\
   SWIFT J1628.1+5145 & Mrk 1498		 &  1920237-1			& 2009-07-31		\\
   SWIFT J1842.0+7945 & 3C 390.3		 &  1340466-1,2,3		& 2008-09-10		\\
   SWIFT J1959.4+4044 & Cygnus A                 &  1420108-1			& 2009-05-09		\\
   SWIFT J2028.5+2543 & MCG+04-48-002            &  1120077-1,2,1122037-1,2	& 2009-05-09,2009-11-09	\\
   SWIFT J2044.2-1045 & Mrk 509		         &  1340467-1,2,3		& 2009-04-30		\\
   SWIFT J2052.0-5704 & IC 5063                  &  1122041-3,4,5		& 2009-10-20,21		\\
   SWIFT J2201.9-3152 & NGC 7172		 &  1122046-1,2,3		& 2009-11-13,14		\\
   SWIFT J2209.4-4711 & NGC 7213                 &  1120069-1,2,3		& 2008-11-08		\\
   SWIFT J2303.3+0852 & NGC 7469                 &  1120055-1,3			& 2008-06-10		\\
   SWIFT J2318.4-4223 & NGC 7582                 &  1122034-2,3,4		& 2009-11-23		\\
\normalsize
\end{longtable}

\newpage
\begin{landscape}
\setlength{\textheight}{480pt}
\setlength{\textwidth}{720pt}
\normalsize
\begin{longtable}{lrrlp{0.8cm}c{0.8cm}ccccr}
   \caption{{\sl Swift}/BAT AGNs with {\sl AKARI}/IRC 2.5--5 $\micron$ Spectra}\label{tab:sample-fluxes}
  \hline      
\hline  
{\sl Swift} Name & R.A. & Dec.& Object Name & Type & $z$ & $\log(L_{\rm 14-195 keV})$\footnotemark[$\ast$] &  $\log(M_{\rm {BH}})$ & $f_{\rm
3.3{\micron}}$ &
$L_{\rm 3.3{\micron}}$ & $N_{\rm H(abs)}$\footnotemark[$\dagger$]\\   
&  &  & & &  & [${\rm erg\,s^{-1}}$] & [$M_{\odot}$]  & [$10^{-14}{\rm erg\,s^{-1}\,cm^{-2}}$] & [$10^{41}\rm erg\,s^{-1}$] &
[$10^{22}\rm cm\,^{-2}$] \\  
\endfirsthead
  \hline\hline

Swift Name & R.A. & Dec.& Object Name & Type & $z$ & $\log(L_{\rm 14-195 keV})$\footnotemark[$\ast$] & $\log(M_{\rm {BH}})$ & $f_{\rm
3.3{\micron}}$ &
$L_{\rm 3.3{\micron}}$ & $N_{\rm H(abs)}$\footnotemark[$\dagger$]\\   
 & &  &  & & & [$\rm erg\,s^{-1}$] & [$M_{\odot}$]  & [$10^{-14}{\rm erg\,s^{-1}\,cm^{-2}}$] & [$10^{41}\rm erg\,s^{-1}$] &
[$10^{22}\rm cm\,^{-2}$] \\  
\hline
 \endhead
  \hline
\endfoot
\hline
\multicolumn{11}{@{}l@{}}{\hbox to 0pt{\parbox{180mm}{\footnotesize \smallskip
            \par\noindent
            \footnotemark[$\ast$] X-ray luminosity in the 14--195 keV band ($\rm erg\,s^{-1}$).
            \par\noindent
            \footnotemark[$\dagger$] $N_{\rm H}$ value taken from \citet{winter09} and \citet{ichikawa12a} and references there in.   
            \par\noindent
            \footnotemark[$\pi$] Classified as ``New Type" object. See \citet{ueda07}
            \par\noindent
            \footnotemark[$\ddagger$] Narrow Line Seyfert 1 (NLS1) object.
          }\hss}}
\endlastfoot
  \hline
   SWIFT J0048.8+3155 &  12.19 &  31.95 & NGC 262         & Sy2.0  & 0.015 &   43.84 & 7.97  &  \textless10.62  &  \textless0.47  & $ 16^{+
4}_{ -3}$ \\
   SWIFT J0134.1-3625 &  23.49 & -36.49 & NGC 612$^{\pi}$    & Sy2.0  & 0.029 &   44.04 & 8.47  &  $17.34^{+1.65}_{-1.85}$ &
$3.07^{+0.29}_{-0.33}$  & $129.7^{+12.9}_{ -8.3}$  \\  
   SWIFT J0138.6-4001 &  24.66 & -40.00 & ESO 297-018$^{\pi}$& Sy2.0  & 0.025 &   44.03 & 9.68  &  $ 2.73^{+1.21}_{-1.20}$ &
$0.34^{+0.15}_{-0.15}$  & $ 41.71^{+ 4.7}_{ -2.9}$  \\  
   SWIFT J0238.2-5213 &  39.56 & -52.20 & ESO 198-024     & Sy1.0  & 0.045 &   44.38 & 8.36  &  \textless 1.45  &  \textless0.61  &   0.100 
                      \\  
   SWIFT J0319.7+4132 &  49.94 &  41.51 & NGC 1275        & Sy2.0  & 0.017 &   43.68 & 8.53  &  \textless20.18  &  \textless1.22  &   0.150 
                      \\
   SWIFT J0554.8+4625 &  88.73 &  46.43 & MCG+08-11-011   & Sy1.5  & 0.020 &   43.96 & 8.07  &  \textless23.87  &  \textless1.98  & $ 
0.250^{+ 0.016}_{ -0.015}$  \\
   SWIFT J0601.9-8636 &  91.47 & -86.60 & ESO 005-G004$^{\pi}$& Sy2.0  & 0.006 &   42.59 & 7.89  &  $2.86^{+1.40}_{-1.42}$ &
$0.02^{+0.01}_{-0.01}$  & $  115$  \\
   SWIFT J0615.8+7101 &  93.93 &  71.02 & Mrk 3           & Sy2.0  & 0.013 &   43.81 & 8.48  &  $1.75^{+0.41}_{-0.41}$ &
$0.06^{+0.01}_{-0.01}$  & $  110$  \\
   SWIFT J0651.9+7426 & 103.04 &  74.42 & Mrk 6$^{\pi}$       & Sy1.5  & 0.018 &   43.78 & 8.24  &  \textless7.27  &   \textless0.51  & $ 
3.26^{+ 1.33}_{ -1.19}$\\
   SWIFT J0902.0+6007 & 135.54 &  60.08 & Mrk 18$^{\pi}$      & Sy2.0  & 0.011 &   42.93 & 7.45  &  $14.11^{+0.95}_{-0.97}$ &
$0.34^{+0.02}_{-0.02}$  & $ 18.25^{+ 3.64}_{ -2.71}$   \\
   SWIFT J0947.6-3057 & 146.92 & -30.94 & MCG-05-23-016   & Sy2.0  & 0.008 &   43.52 & 7.66  &  \textless7.22  &   \textless0.10  & $ 
1.600^{+ 0.005}_{ -0.006}$  \\
   SWIFT J0959.5-2248 & 149.86 & -22.82 & NGC 3081$^{\pi}$    & Sy2.0  & 0.007 &   43.16 & 7.96  &  $3.80^{+1.17}_{-1.19}$  &
$0.05^{+0.01}_{-0.01}$  & $ 94.2^{+ 6.2}_{ -7.2}$  \\
   SWIFT J1031.7-3451 & 157.95 & -34.86 & NGC 3281$^{\pi}$    & Sy2.0  & 0.010 &   43.36 & 8.62  &  \textless10.07  &  \textless0.22  & $
86.30^{+16.32}_{-16.12}$  \\
   SWIFT J1049.4+2258 & 162.38 &  22.97 & Mrk 417$^{\pi}$     & Sy2.0  & 0.032 &   43.97 & 8.04  &  \textless 3.44  &  \textless0.74  & $
85.69^{+12.73}_{ -6.96}$ \\
   SWIFT J1106.5+7234 & 166.68 &  72.57 & NGC 3516        & Sy1.5  & 0.008 &   43.34 & 8.13  &  \textless 4.85  &  \textless0.07  & $ 
0.353^{+ 0.32}_{ -0.12}$ \\
   SWIFT J1143.7+7942 & 176.15 &  79.67 & UGC 06728       & Sy1.2  & 0.006 &   42.44 & 6.81  &  \textless 1.15  &  \textless0.01  & $ 
0.01^{+ 0.01}_{ -0.01}$  \\
   SWIFT J1206.2+5243 & 181.59 &  52.72 & NGC 4102        & Sy2.0  & 0.002 &   41.66 & 7.90  &  $134.66^{+7.16}_{-7.28}$ &
$0.20^{+0.01}_{-0.01}$  & $200$  \\
   SWIFT J1210.5+3924 & 182.63 &  39.40 & NGC 4151        & Sy1.5  & 0.003 &   43.18 & 7.69  &  $13.00^{+8.66}_{-8.72}$ &
$0.03^{+0.02}_{-0.02}$  & $  5.32^{+ 0.07}_{ -0.08}$  \\
   SWIFT J1225.8+1240 & 186.44 &  12.66 & NGC 4388        & Sy2.0  & 0.008 &   43.74 & 8.53  &  $8.14^{+1.87}_{-1.88}$ &
$0.11^{+0.03}_{-0.03}$  & $ 36.17^{+ 3.81}_{ -3.82}$  \\
   SWIFT J1238.9-2720 & 189.73 & -27.30 & ESO 506-G027$^{\pi}$& Sy2.0  & 0.025 &   44.29 & 8.59  &  $2.81^{+1.08}_{-1.09}$ &
$0.35^{+0.14}_{-0.14}$  & $ 76.82^{+ 7.37}_{ -6.79}$  \\
   SWIFT J1322.2-1641 & 200.62 & -16.74 & MCG-03-34-064   & Sy1.8  & 0.016 &   43.29 & 8.28  &  $4.54^{+1.48}_{-1.49}$ &
$0.24^{+0.08}_{-0.08}$  & $ 40.73^{+ 4.79}_{ -4.30}$  \\
   SWIFT J1338.2+0433 & 204.57 &   4.54 & NGC 5252        & Sy1.9  & 0.022 &   43.99 & 8.64  &  \textless 3.94  &  \textless0.41  & $ 
4.34^{+ 0.52}_{ -0.42}$  \\
   SWIFT J1413.2-0312 & 213.30 &  -3.20 & NGC 5506        & Sy1.9  & 0.006 &   43.34 & 7.77  &  $15.25^{+5.81}_{-5.85}$ &
$0.11^{+0.04}_{-0.04}$ & $  2.78^{+ 0.05}_{ -0.05}$  \\
   SWIFT J1442.5-1715 & 220.60 & -17.23 & NGC 5728        & Sy2.0  & 0.009 &   43.31 & 8.53  &  $18.95^{+5.41}_{-4.83}$ &
$0.32^{+0.09}_{-0.08}$ & $ 82.0^{+ 5.3}_{ -5.0}$  \\
   SWIFT J1959.4+4044 & 299.89 &  40.73 & Cygnus A        & Sy2.0  & 0.056 &   44.96 & 9.39  &  \textless 8.03  &  \textless5.25  & $
11^{+21}_{ -6}$ \\
   SWIFT J2028.5+2543 & 307.14 &  25.73 & MCG+04-48-002   & Sy2.0  & 0.013 &   43.58 & 7.50  &  $49.46^{+2.15}_{-2.19}$ &
$1.87^{+0.08}_{-0.08}$  & $ 96.00^{+51.97}_{-27.77}$  \\
   SWIFT J2052.0-5704 & 313.00 & -57.07 & IC 5063$^{\pi}$     & Sy2.0  & 0.011 &   43.39 & 7.68  &  $8.00^{+2.62}_{-2.66}$ &
$0.20^{+0.07}_{-0.07}$  & $ 21.78^{+ 2.24}_{ -2.06}$  \\
   SWIFT J2209.4-4711 & 332.32 & -47.16 & NGC 7213        & Sy1.5  & 0.005 &   42.64 & 8.63  &  \textless 6.30  &  \textless0.04  & $ 
0.025^{+ 0.011}_{ -0.012}$ \\
   SWIFT J2303.3+0852 & 345.81 &   8.86 & NGC 7469        & Sy1.2  & 0.016 &   43.60 & 8.64  &  $79.76^{+6.73}_{-7.11}$ &
$4.16^{+0.35}_{-0.37}$  &   0.041                        \\
   SWIFT J2318.4-4223 & 349.59 & -42.36 & NGC 7582        & Sy2.0  & 0.005 &   42.68 & 8.31  &  $84.26^{+6.77}_{-6.62}$ &
$0.45^{+0.04}_{-0.04}$  & $  33$  \\
   SWIFT J0623.9-6058 &  95.98 & -60.97 & ESO 121-G028	  & Sy2.0  & 0.040 &   44.03 & 9.00  &  \textless 0.81  &  \textless0.27  & $
16.19^{+12.6}_{- 9.4}$ \\
   SWIFT J0920.8-0805 & 140.21 &  -8.07 & MCG-01-24-012	  & Sy2.0  & 0.019 &   43.60 & 7.16  &  \textless 6.59  &  \textless0.50  & $
11.44^{+ 2.82}_{ -2.27}$ \\
   SWIFT J1628.1+5145 & 247.04 &  51.75 & Mrk 1498$^{\pi}$    & Sy1.9  & 0.054 &   44.49 & 8.59  &  \textless10.18  &  \textless6.32  & $
17.84^{+ 2.37}_{ -1.82}$ \\
   SWIFT J0123.9-5846 &  20.94 & -58.79 & Fairall 9	  & Sy1.2  & 0.047 &   44.42 & 8.91  &  \textless 3.84  &  \textless1.74  &    0.023
                      \\
   SWIFT J2044.2-1045 &  33.69 &  -0.79 & Mrk 590	  & Sy1.0  & 0.026 &   43.43 & 8.87  &  \textless 4.13  &  \textless0.57  &   0.027 
                      \\
   SWIFT J0426.2-5711 &  66.50 & -57.20 & 1H 0419-577	  & Sy1.5  & 0.104 &   44.77 & 9.00  &  \textless 5.63  &  \textless13.53  & $204$  
  \\
   SWIFT J0433.0+0521 &  68.29 &   5.36 & 3C 120	  & Sy1.0  & 0.033 &   44.48 & 8.56  &  $5.44^{+2.62 }_{-2.66}$ &
$1.19^{+0.57}_{-0.58}$  & $  0.16^{+ 0.01}_{ -0.01}$  \\
   SWIFT J0516.2-0009 &  79.05 &  -0.15 & Ark 120	  & Sy1.0  & 0.032 &   44.23 & 8.74  &  \textless 6.76  &  \textless1.45  &   0.020 
                      \\
   SWIFT J0519.5-3140 &  80.74 & -36.45 & ESO 362-G021	  & Sy1.0  & 0.056 &   42.27 & 9.00  &  \textless 2.63  &  \textless1.75  &   0.010 
                      \\
   SWIFT J0742.5+4948 & 115.60 &  49.81 & Mrk 79	  & Sy1.2  & 0.022 &   43.74 & 8.42  &  \textless 7.42  &  \textless0.72  &   0.006 
                     \\
   SWIFT J0925.0+5218 & 141.30 &  52.28 & Mrk 110$^{\ddagger}$	  & Sy1.0  & 0.035 &   44.25 & 7.80  &  \textless 6.20  &  \textless1.56  &
$  0.02^{+ 0.01}_{ -0.01}$ \\
   SWIFT J0945.6-1420 & 146.44 & -14.32 & NGC 2992	  & Sy1.9  & 0.007 &   42.80 & 8.04  &  $8.88^{+1.75}_{-1.77}$ &
$0.10^{+0.02}_{-0.02}$  & $  1.19^{+ 2.21}_{ -0.09}$  \\
   SWIFT J1139.0-3743 & 174.76 & -37.74 & NGC 3783	  & Sy1.5  & 0.009 &   43.61 & 8.21  &  \textless12.29  &  \textless0.23  & $ 
0.57^{+ 0.21}_{ -0.14}$ \\
   SWIFT J1203.0+4433 & 180.78 &  44.52 & NGC 4051$^{\ddagger}$	  & Sy1.5  & 0.002 &   41.71 & 7.27  &  \textless18.76  &  \textless0.02  & 
 0.029                       \\
   SWIFT J1239.6-0519 & 189.91 &  -5.34 & NGC 4593	  & Sy1.0  & 0.009 &   43.25 & 8.61  &  \textless 9.74  &  \textless0.15  & $ 
0.031^{+ 0.011}_{ -0.012}$ \\
   SWIFT J1303.8+5345 & 196.02 &  53.78 & SBS 1301+540	  & Sy1.0  & 0.029 &   43.92 & 7.54  &  \textless 3.37  &  \textless0.60  &   0.060 
                      \\
   SWIFT J1305.4-4928 & 196.36 & -49.46 & NGC 4945	  & Sy2.0  & 0.001 &   42.41 & 6.04  &  $273.00^{+23.51}_{-24.31}$ &
$0.18^{+0.02}_{-0.02}$  & $ 530$  \\
   SWIFT J1349.3-3018 & 207.32 & -30.30 & IC 4329A	  & Sy1.2  & 0.016 &   44.28 & 8.52  &  $16.81^{+5.52}_{-5.54}$ &
$0.85^{+0.28}_{-0.28}$  & $  0.61^{+ 0.03}_{ -0.03}$  \\
   SWIFT J1352.8+6917 & 208.26 &  69.30 & Mrk 279	  & Sy1.5  & 0.030 &   44.05 & 8.62  &  \textless 5.19  &  \textless0.96  &   0.013 
                      \\
   SWIFT J1417.9+2507 & 214.49 &  25.13 & NGC 5548	  & Sy1.5  & 0.017 &   43.73 & 8.42  &  $ 4.21^{+2.12}_{-2.14}$ &
$0.24^{+0.12}_{-0.12}$ & $  0.07^{+ 0.04}_{ -0.05}$  \\
   SWIFT J1535.9+5751 & 233.97 &  57.87 & Mrk 290	  & Sy1.5  & 0.029 &   43.71 & 7.68  &  \textless 2.68  &  \textless0.47  & $ 
0.15^{+ 0.03}_{ -0.05}$ \\
   SWIFT J1842.0+7945 & 280.55 &  79.77 & 3C 390.3	  & Sy1.0  & 0.056 &   44.92 & 8.52  &  \textless 3.40  &  \textless2.22  & $ 
0.12^{+ 0.03}_{ -0.03}$ \\
   SWIFT J2044.2-1045 & 311.03 & -10.72 & Mrk 509	  & Sy1.5  & 0.034 &   44.41 & 8.59  &  $8.91^{+2.87}_{-2.89}$ &
$2.12^{+0.68}_{-0.69}$  & $  0.015^{+ 0.008}_{-0.008}$  \\
   SWIFT J2201.9-3152 & 330.51 & -31.86 & NGC 7172	  & Sy2.0  & 0.008 &   43.48 & 8.31  & $26.59^{+3.92}_{-3.96}$ &
$0.39^{+0.06}_{-0.06}$   & $  8.19^{+ 3.42}_{ -3.30}$  \\
\setlength{\textheight}{680pt}
\setlength{\textwidth}{512pt}
\normalsize
\end{longtable}
\end{landscape}

\newpage
\begin{longtable}{lp{1cm}p{1cm}cccr}
  \caption{Linear regression parameters obtained using the E-M method under ASURV.}\label{tab:regressions}
  \hline    
  \hline    

\endfirsthead
  \hline
\endhead
  \hline
\endfoot
  \hline

\multicolumn{7}{@{}l@{}}{\hbox to 0pt{\parbox{180mm}{\footnotesize \smallskip
            \par\noindent
            \footnotemark[$\ast$] Number of upper-limits in the subsample.
            \par\noindent
            \footnotemark[$\dagger$] Correlation probability by Cox's proportional hazard model.  

          }\hss}}
\endlastfoot

\multicolumn{7}{c}{$\log(L_{\rm 3.3{\micron}}) =  a_{0}\;\;  \{ \log(L_{\rm 14-195keV})-c_{0}\} \;+\; b_{0}$}\\ 
Sample & No. & Up.\footnotemark[$\ast$] & a$_0$ & b$_0$ & $\langle\log(L_{\rm 3.3{\micron}})\rangle$ &  P\footnotemark[$\dagger$] \\ 
\hline
 All AGNs      & 54 & 30 & 0.42$\pm$0.14  & 40.20$\pm$0.11 & 40.07$\pm$0.12 & 0.01 \\
 Optical type 1& 26 & 20 & 1.05$\pm$0.27  & 39.92$\pm$0.24 & 39.68$\pm$0.22 & 0.02 \\
 Optical type 2& 28 & 10 & 0.11$\pm$0.17  & 40.27$\pm$0.12 & 40.23$\pm$0.11 & 0.66 \\ 
 X-ray type 1  & 24 & 19 & 1.11$\pm$0.35  & 39.83$\pm$0.33 & 39.61$\pm$0.24 & 0.05 \\
 X-ray type 2  & 30 & 11 & 0.12$\pm$0.18  & 40.23$\pm$0.12 & 40.19$\pm$0.11 & 0.56 \\
\hline 
\hline
\multicolumn{7}{c}{$\log(L_{\rm 3.3{\micron}}/M_{\rm {BH}}) =  a_{1}\;\;  \{ \log(L_{\rm 14-195keV}/M_{\rm {BH}})-c_{1}\} \;+\; b_{1}$}\\   
  
Sample & No. & Up. & a$_1$ & b$_1$ & $\langle\log(L_{\rm 3.3{\micron}}/M_{\rm {BH}})\rangle$ &  P\\ 
\hline
 All  AGNs     & 54 & 30 & 0.73$\pm$0.17 & 31.97$\pm$0.13 & 31.88$\pm$0.13 & 0.002 \\
 Optical type 1& 26 & 20 & 1.56$\pm$0.37 & 31.55$\pm$0.26 & 31.62$\pm$0.21 & 0.005 \\
 Optical type 2& 28 & 10 & 0.62$\pm$0.23 & 32.16$\pm$0.16 & 32.01$\pm$0.16 & 0.10 \\
 X-ray type 1  & 24 & 19 & 1.59$\pm$0.43 & 31.48$\pm$0.33 & 31.53$\pm$0.21 & 0.006 \\
 X-ray type 2  & 30 & 11 & 0.58$\pm$0.22 & 32.13$\pm$0.15 & 32.00$\pm$0.15 & 0.10 \\
\hline
\hline
\multicolumn{7}{c}{$\log(L_{\rm 3.3{\micron}})  =   a_{2}\;\;  \{ \log(N_{H})-  c_{2}\}  \;+\; b_{2}$}\\         
Sample & No. & Up. & a$_2$ & b$_2$ & $\langle\log(L_{\rm 3.3{\micron}})\rangle$ &  P\\ 
\hline
 All           & 54 & 30 & 0.14$\pm$0.09 & 40.02$\pm$0.14 & 40.07$\pm$0.12 & 0.003 \\
$\log(L_{\rm 14-195keV})\leq 43.64$ & 26 & 11 &  0.28$\pm$0.13 & 39.78$\pm$0.19 & 39.95$\pm$0.15 & 0.01 \\
$\log(L_{\rm 14-195keV})>    43.64$ & 28 & 19 & -0.03$\pm$0.09 & 40.39$\pm$0.15 & 40.37$\pm$0.13 & 0.58 \\
\hline
\hline
\multicolumn{7}{c}{$\log(L_{\rm 3.3{\micron}}/M_{\rm {BH}}) =   a_{3}\;\;  \{ \log(N_{H})-  c_{3}\}  \;+\; b_{3}$}\\
Sample & No. & Up. & a$_3$ & b$_3$ & $\langle\log(L_{\rm 3.3{\micron}}/M_{\rm {BH}})\rangle$ &  P \\ 
\hline
 All           & 54 & 30 & 0.22$\pm$0.12 & 31.71$\pm$0.19 & 31.88$\pm$0.13 & 0.075 \\
$\log (L_{\rm 14-195keV}/M_{\rm BH})\leq 35.43$ & 25 & 13 & 0.15$\pm$0.10 & 31.44$\pm$0.15 & 31.52$\pm$0.12 & 0.12 \\
$\log (L_{\rm 14-195keV}/M_{\rm BH})>    35.43$ & 29 & 17 & 0.20$\pm$0.14 & 32.23$\pm$0.23 & 32.42$\pm$0.14 & 0.17 \\
\end{longtable}
\begin{longtable}{llp{0.6cm}ccccccc}
  \caption{Two sample tests for optically and X-ray classified AGNs.}\label{tab:twost-results}
  \hline  
  \hline      
Class.& Criteria &  & & & \multicolumn{2}{c}{$\langle\log(L_{\rm 3.3{\micron}}/M_{\rm {BH}})\rangle$}&  Gehan's & logrank & Peto\&Peto \\
 &   &  No. &$n_{1}$\footnotemark[$\ast$]& $n_{2}$\footnotemark[$\dagger$]&Type 1 & Type 2 & Prob. & Prob. & Prob.\\
\endfirsthead
  \hline
\endhead
  \hline
\endfoot
  \hline
\multicolumn{11}{@{}l@{}}{\hbox to 0pt{\parbox{180mm}{\footnotesize \smallskip
            \par\noindent
            \footnotemark[$\ast$] Number of type 1 objects contained in the sample.
            \par\noindent
            \footnotemark[$\dagger$] Number of type 2 objects contained in the sample.

          }\hss}}
\endlastfoot
  \hline
Optical & All AGNs                      & 54 & 26 & 28 & 31.62$\pm$0.21 & 32.01$\pm$0.16 & 0.27 & 0.17 & 0.21 \\
& $\log(L_{X}/M_{\rm {BH}}) \leq 35.43$ & 25 & 10 & 15 & 31.17$\pm$0.15 & 31.66$\pm$0.13 & 0.14 & 0.09 & 0.12\\
& $\log(L_{X}/M_{\rm {BH}}) > 35.43$    & 29 & 16 & 13 & 32.21$\pm$0.19 & 32.62$\pm$0.20 & 0.21 & 0.18 & 0.18\\
X-ray   & All AGNs                      & 54 & 24 & 30 & 31.53$\pm$0.21 & 32.00$\pm$0.15 & 0.34 & 0.14 & 0.20 \\
\hline
\hline
Class.& Criteria & & & & \multicolumn{2}{c}{$\langle\log(L_{\rm 3.3{\micron}})\rangle$}&  Gehan's & logrank & Peto\&Peto \\
 &   & No.  &$n_{1}$ & $n_{2}$&Type 1 & Type 2 & Prob. & Prob. & Prob.\\
\hline
Optical & All AGNs         & 54 & 26 & 28 & 39.68$\pm$0.22 & 40.23$\pm$0.11 & 0.58 & 0.05 & 0.22 \\
& $\log(L_{X}) \leq 43.64$ & 26 &  9 & 17 & 39.44$\pm$0.28 & 40.19$\pm$0.13 & 0.04 & 0.02 & 0.02\\
& $\log(L_{X}) > 43.64$    & 28 & 17 & 11 & 40.59$\pm$0.09 & 40.29$\pm$0.19 & 0.66 & 0.34 & 0.50\\ 
X-ray   & All AGNs         & 54 & 24 & 30 & 39.61$\pm$0.24 & 40.19$\pm$0.11 & 0.80 & 0.08 & 0.37 \\
\end{longtable}

\newpage

\newpage
\begin{figure}
  \begin{center}
    \FigureFile(160mm,80mm){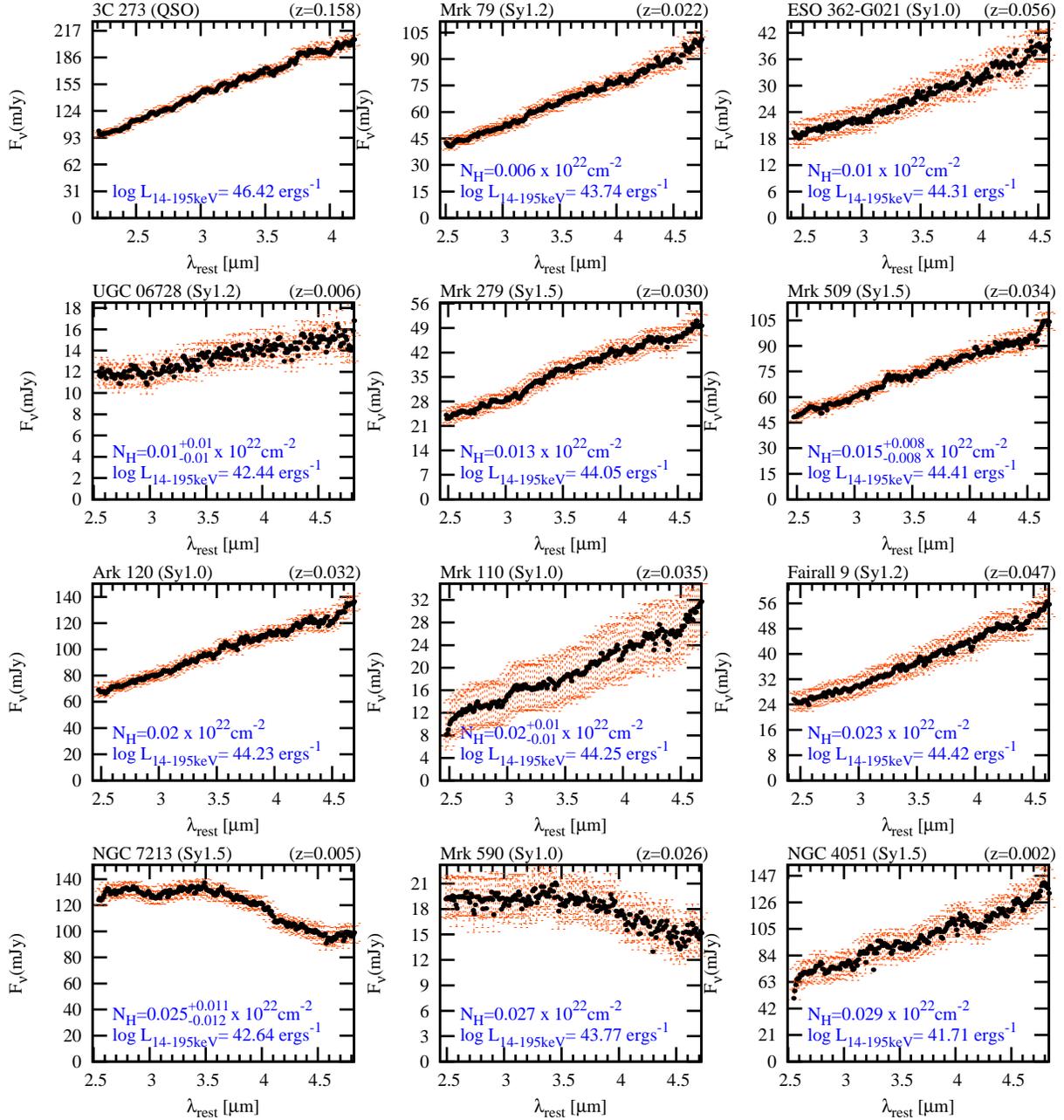}
  \caption{{\sl AKARI}/IRC infrared 2.5--5 $\micron$ spectra of our X-ray selected AGN sample. The abscissa is the rest-frame
wavelength and the ordinate is the flux $F_{\nu}$ in mJy. Each object shows its name, redshift and $N_{\rm H}$ column density values. X-ray
luminosities (in units of ${\rm erg\,s^{-1}}$) in the band 14--195 keV ($L_{14-195keV}$) were obtained from \citet{tueller08}.
(Color figures are available on the electronic version only.)}
  \end{center}
\end{figure}
\begin{figure}
  \begin{center}
    \ContinuedFloat
    \FigureFile(160mm,80mm){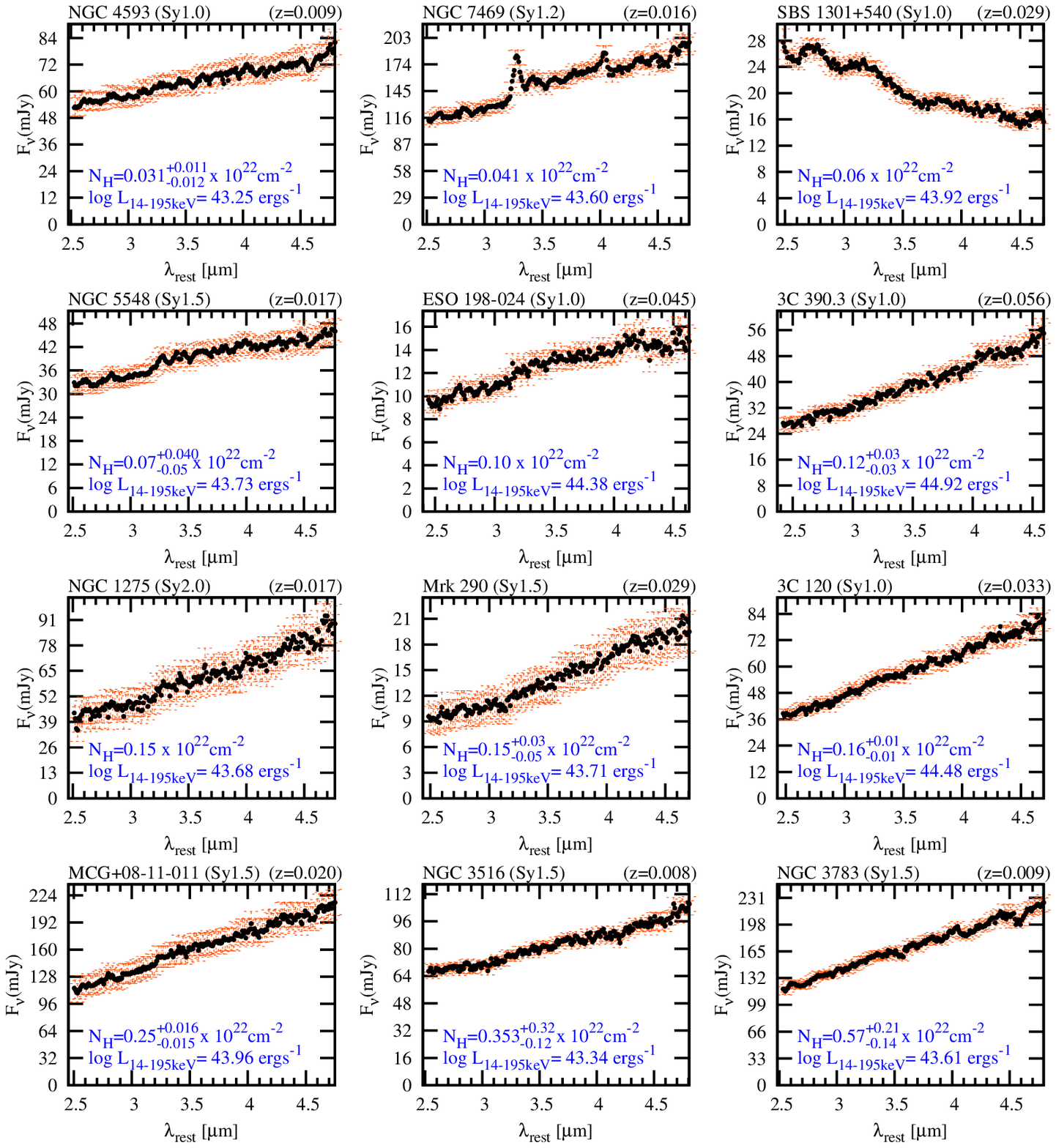}
  \end{center}
\caption{Continued.}
\end{figure}
\begin{figure}
  \begin{center}
    \ContinuedFloat
    \FigureFile(160mm,80mm){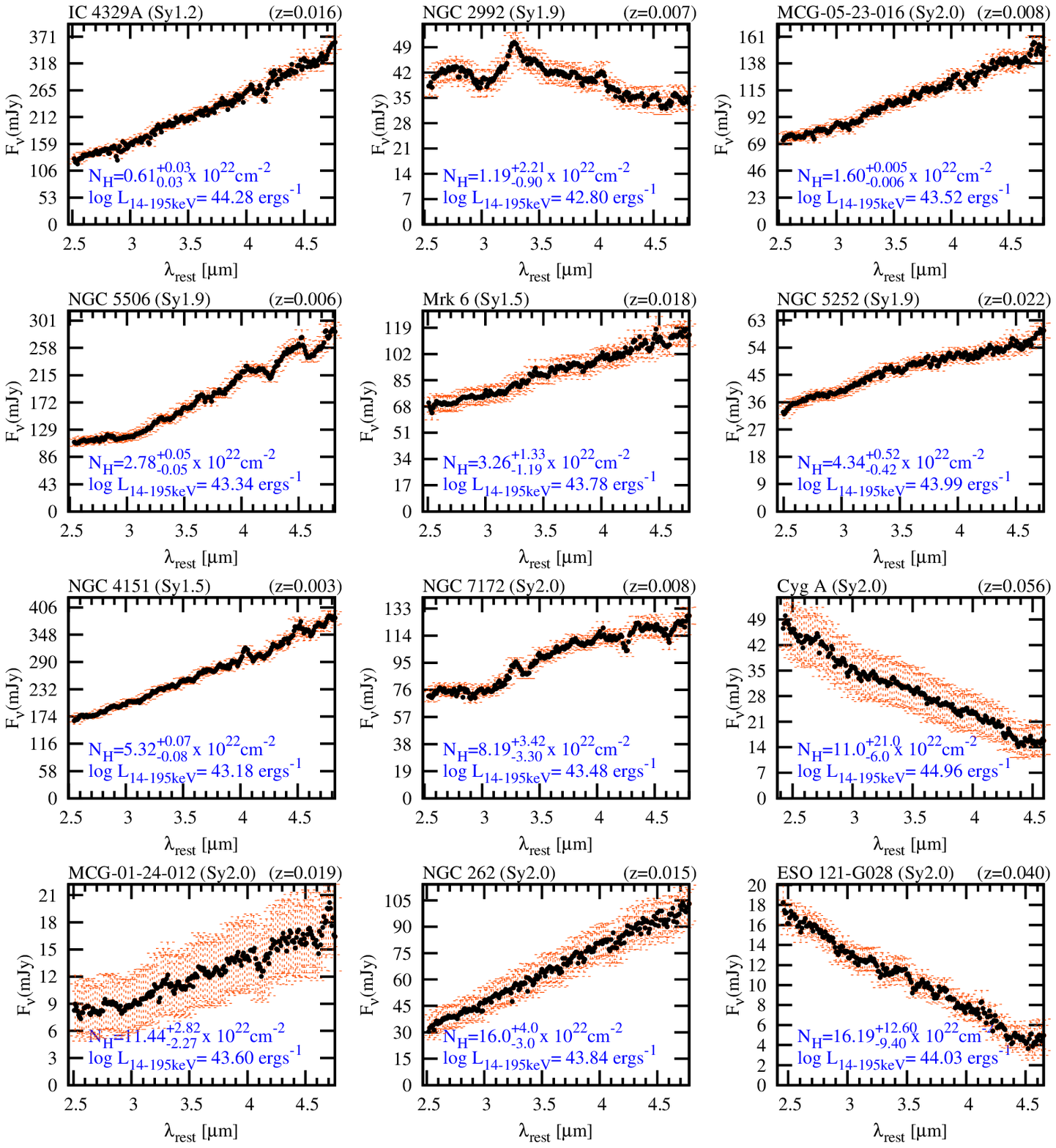}
  \end{center}
  \caption{Continued.}
\end{figure}
\begin{figure}
  \begin{center}
    \ContinuedFloat
    \FigureFile(160mm,80mm){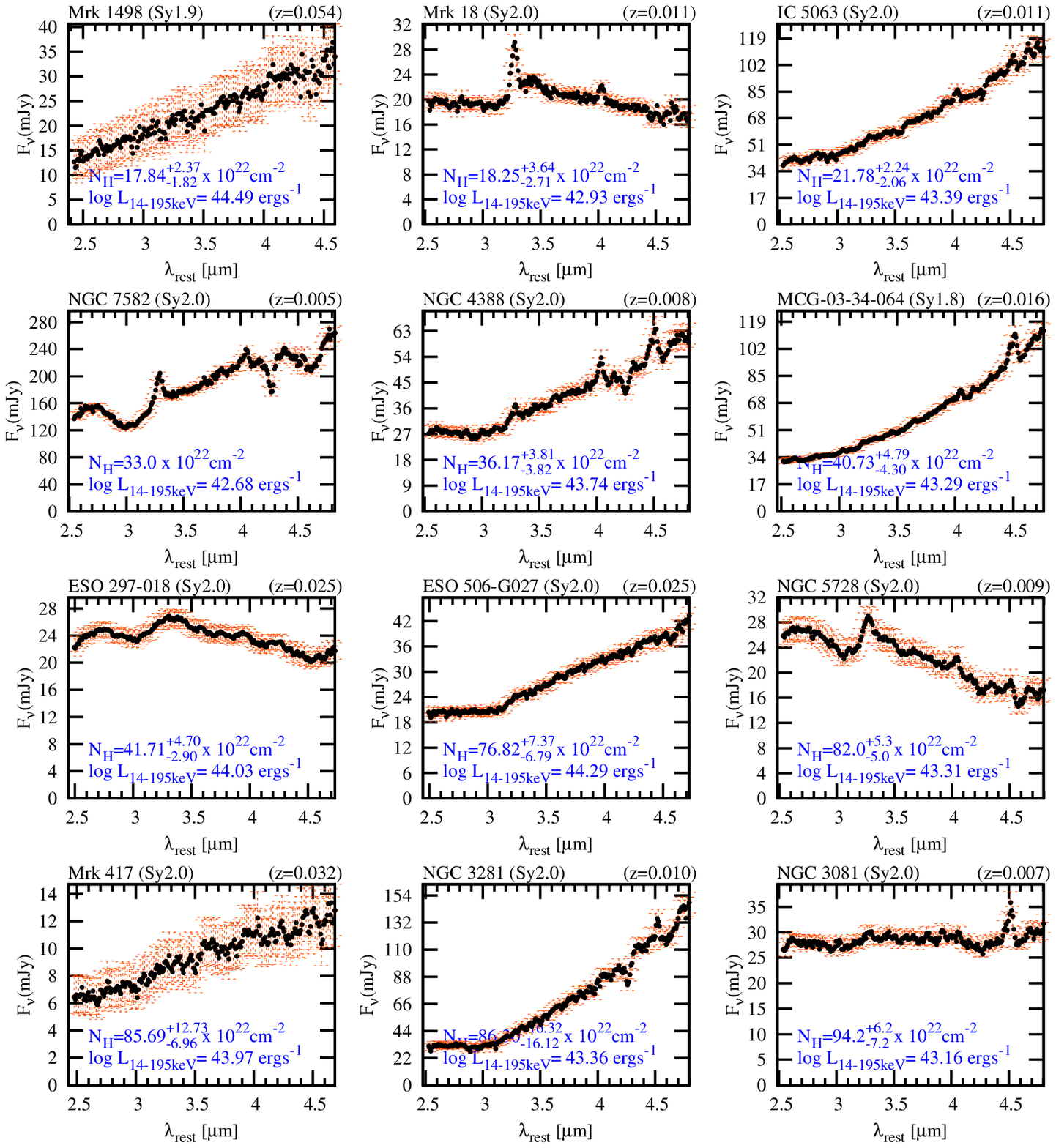}
  \end{center}
  \caption{Continued.}
\end{figure}
\begin{figure}
  \begin{center}
    \ContinuedFloat
    \FigureFile(160mm,80mm){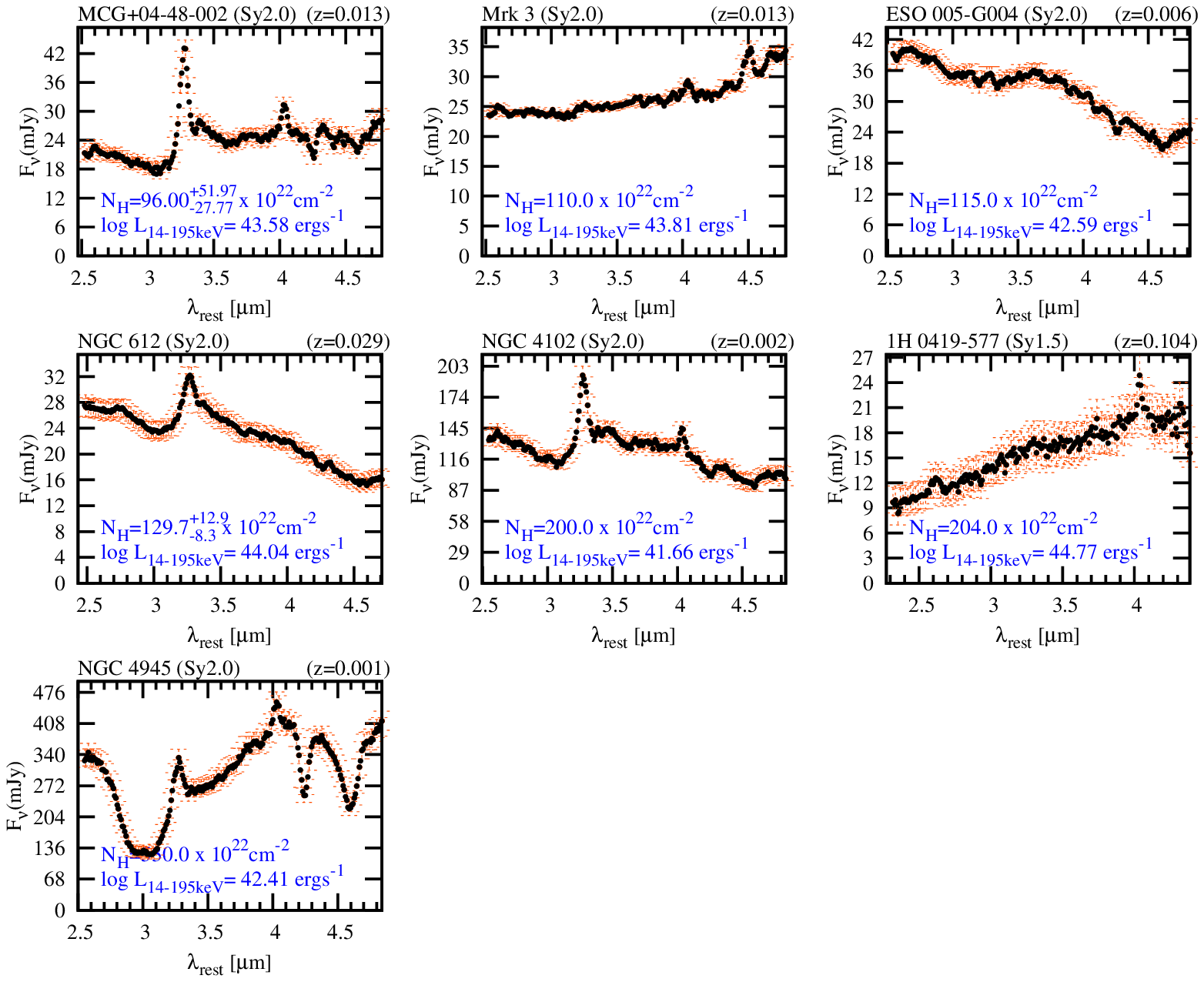}
  \end{center}
  \caption{Continued.}
  \label{fig:all-spectra}
\end{figure}

\newpage
\newpage

\newpage

\newpage

\newpage

\end{document}